\documentclass[3p,times]{elsarticle}
\usepackage{mathtools}
\usepackage{siunitx}
\usepackage{xcolor}
\usepackage[numbers]{natbib} 
\usepackage[utf8]{inputenc}
\usepackage[english]{babel}
\usepackage[colorlinks,allcolors=blue]{hyperref}
\usepackage{todonotes} 
\usepackage{caption}

\newcommand{\e}[1]{\times 10^{#1}}
\usepackage{subcaption,siunitx,booktabs}
\usepackage{subcaption}
\usepackage{colortbl}
\usepackage[english]{babel} 
\usepackage{amsmath,amsfonts,amsthm,bm}
\usepackage{caption}
\usepackage{float}

\usepackage{setspace}

\begin{document}
\doublespacing
\begin{frontmatter}

\title{Optimising cryptocurrency portfolios through stable clustering of price correlation networks}

\author[1]{Ruixue Jing}
\author[2,3]{Ryota Kobayashi}
\author[1,4]{Luis E. C. Rocha}
\address[1]{Department of Economics, Ghent University, Ghent, 9000, Belgium}
\address[2]{Graduate School of Frontier Sciences, The University of Tokyo, Kashiwa, 277-8561, Japan}
\address[3]{Mathematics and Informatics Center, The University of Tokyo, Tokyo, 113-8656, Japan}
\address[4]{Department of Physics and Astronomy, Ghent University, Ghent, 9000, Belgium}

\begin{abstract}
The rapidly evolving cryptocurrency market presents unique challenges for investment due to its inherent volatility and evolving regulatory environment. Collective price movements can be exploited to construct diversified portfolios with improved risk–return profiles. This paper introduces an integrated framework that combines network analysis, price forecasting, and portfolio theory to identify stable groups of highly correlated cryptocurrencies for profitable portfolio construction. We employ the Louvain community detection algorithm together with consensus clustering to extract temporally persistent correlation clusters, and incorporate ARIMA-based price forecasts to enhance forward-looking cluster formation. Using 5 years of daily closing prices, we evaluate portfolio performance across multiple strategies and holding horizons, assessing both profitability and downside risk with return-based and tail-risk metrics. Our empirical results show that predictive consensus-clustering portfolios maintain consistently positive and stable performance up to a 14-day horizon, exhibit favourable gain–loss asymmetry, and achieve tighter tail-risk control. These findings demonstrate that stable interdependencies in cryptocurrency markets can be leveraged to construct profitable and risk-aware portfolios across short-term holding horizons.
\end{abstract}

\begin{keyword}
Return forecasting, Correlation networks, Consensus clustering, Portfolio optimisation, Tail-risk management
\end{keyword}

\end{frontmatter}

\section{Introduction}

The global cryptocurrency market has experienced rapid expansion and increasing financial relevance over the past decade. Since the introduction of Bitcoin in 2009~\cite{elbahrawy2017evolutionary}, cryptocurrencies have evolved from a niche technological experiment into a recognised asset class~\cite{legotin2018prospects, antonakakis2019cryptocurrency}. Their decentralised nature enables peer-to-peer transactions, reduces reliance on intermediaries, and offers new avenues for diversification, high-growth opportunities, and financial inclusion~\cite{corbet2019cryptocurrencies}. The success of Bitcoin spurred the creation of thousands of alternative coins with heterogeneous technological features and market behaviours~\cite{arslanian2022emergence}. 
At the same time, the market remains challenged by extreme volatility, regulatory uncertainty, and security risks such as hacking and fraud~\cite{arsi2022cryptocurrencies, msefula2024financial}. 
A further structural concern is the susceptibility of cryptocurrency markets to price manipulation~\cite{li2018cryptocurrency}, which can distort short-term price formation and exacerbate uncertainty for both forecasting models and portfolio construction. Despite these limitations, institutional adoption continues to grow, reinforcing cryptocurrencies' evolving role within global finance. Understanding their potential benefits and risks is essential for investors navigating this dynamic market.

Predicting cryptocurrency price movements is difficult due to large fluctuations, non-stationarity, and complex market dynamics~\cite{pintelas2020investigating}. Yang et al.~\cite{yang2019price} show that cryptocurrencies exhibit higher entropy and conditional entropy than traditional equities, implying elevated unpredictability and the limited robustness of conventional time-series models across regimes. This has motivated the use of diverse predictive approaches, including ARIMA and LSTM models~\cite{khedr2021cryptocurrency}, deep learning architectures tailored for trading and risk management~\cite{patel2020deep, vikram2022crypto}, and technical-rule-based strategies that mitigate downside risk~\cite{deprez2024simple}. These studies highlight the need for adaptive prediction methods capable of capturing regime shifts and rapid price fluctuations. In this context, portfolio construction becomes a central tool for managing cryptocurrency risk.
Within traditional portfolio theory, diversification and asset allocation provide fundamental tools for balancing risk and return~\cite{goetzmann2008equity, basak2007optimal}.
Modern Portfolio Theory (MPT) provides a foundational framework for balancing return and risk and has been applied successfully in cryptocurrency markets, where MPT-based portfolios can outperform individual assets~\cite{brauneis2019cryptocurrency, fabozzi2008portfolio}. Beyond MPT, risk-parity allocations~\cite{chaves2011risk}, algorithmic trading, and machine learning~\cite{amirzadeh2022applying}, including reinforcement learning~\cite{jiang2017cryptocurrency}, offer additional tools for constructing adaptive, data-driven portfolios. In all these cases, the dependence structure among cryptocurrencies plays a central role in determining diversification benefits.

Portfolio formation in cryptocurrencies has also embraced alternative data sources and clustering-based approaches. Sentiment analysis from social media or news~\cite{karalevicius2018using, kim2022dynamics} and machine-learning-based classification of assets~\cite{jiang2017cryptocurrency, rao2021crypto} provide insights into market structure beyond prices. Blockchain analytics contributes on-chain information regarding liquidity, major wallet movements, and abnormal transaction patterns~\cite{zuniga2023maximizing, lan2025risk}, enabling more informed allocation decisions. 
A central component of diversification is the correlation structure between assets. Network modelling provides a powerful framework for analysing these relationships, revealing complex interdependencies beyond simple pairwise correlations~\cite{hong2022impact}. Network-based portfolio methods, such as hierarchical clustering trees and Minimum Spanning Trees (MSTs) have been shown to extract meaningful structure in stock markets and enhance diversification~\cite{rea2014visualization}. In dynamic settings, rolling-window and MST-based networks highlight low-correlation assets positioned at the network periphery ~\cite{lyocsa2012stock, onnela2003dynamics, jing2023network}, but MSTs require transforming correlations into distances and pruning edges, which may eliminate meaningful structural information~\cite{ioannidis2023portfolio}.
Planar Maximally Filtered Graphs (PMFGs), in contrast, retain a larger set of strong correlations while enforcing planarity, providing a richer representation of dependency structure for portfolio selection~\cite{tumminello2005tool}.
Assessing downside risk is equally essential in cryptocurrency markets, which exhibit heavy tails and systemic co-movement. Standard deviation alone fails to capture extreme losses, motivating the use of risk measures such as Value-at-Risk (VaR), Marginal Expected Shortfall (MES), and the Omega ratio~\cite{mcneil2015quantitative}.
These metrics complement simple profitability indicators, yielding a more comprehensive understanding of portfolio resilience, which is particularly relevant for highly volatile cryptocurrency portfolios~\cite{kaufman2013trading,pardo2011evaluation}.

Identifying stable clusters of highly correlated cryptocurrencies is therefore crucial for effective diversification. Community detection methods such as the Louvain algorithm~\cite{blondel2008fast} reveal densely connected groups of assets, providing insights into the underlying market structure. Yet cryptocurrency networks are highly dynamic, and community partitions may fluctuate across time or algorithmic runs due to their stochastic nature. This transience can obscure stable intrinsic interdependencies needed for robust portfolio construction. Consensus clustering~\cite{lancichinetti2012consensus} offers a principled solution by aggregating multiple partitions into a stable consensus structure. In this study, we present a consensus-based clustering framework specifically designed to extract persistent clusters from dynamic cryptocurrency correlation networks. By integrating predictive time-series modelling with network-based community detection and portfolio theory, our approach identifies enduring interdependencies, enhances the stability of cluster detection, and supports the construction of diversified portfolios that balance historical insights with predictive accuracy.

\section{Materials and Methods}

\begin{figure}[h]
    \centering
    \includegraphics[scale=0.72]{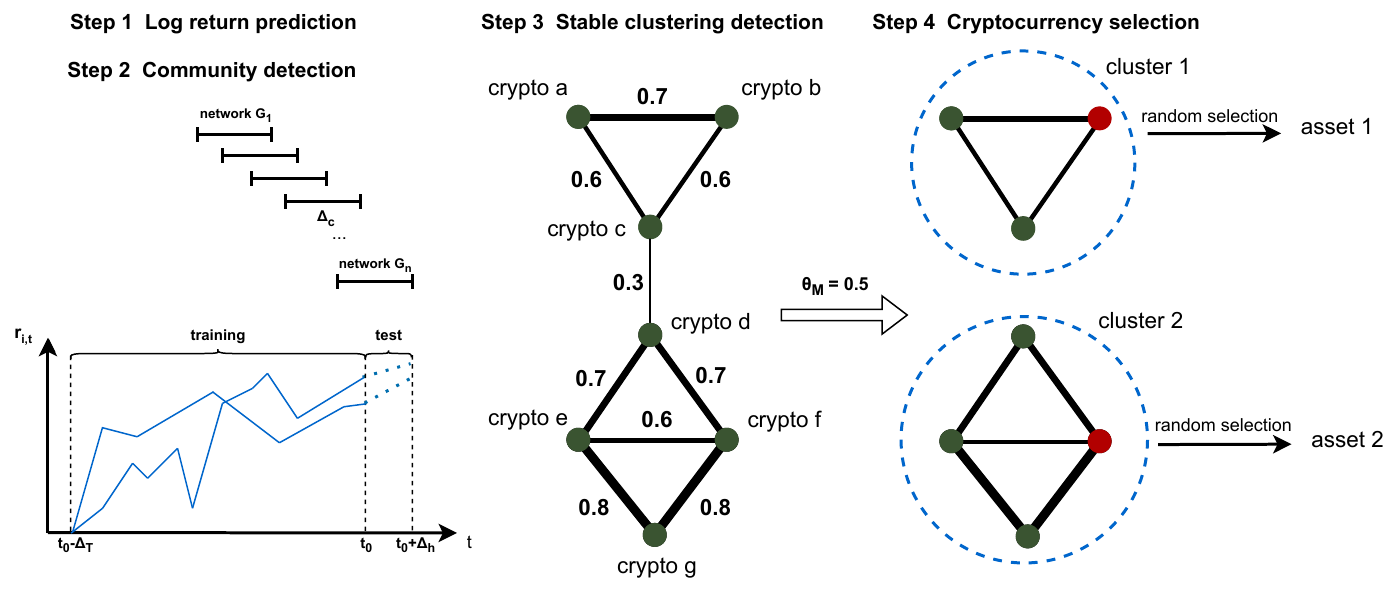}
    \caption{The workflow of our portfolio construction method, including price log return prediction (step 1), community detection (step 2), stable clustering detection (step 3), and cryptocurrency selection (step 4). The full blue lines in step 1 represent the empirical price log return data of the cryptocurrencies, while the dashed blue lines indicate the predicted value. The $n$ networks (obtained in step 2) are aggregated in step 3. The blue circles in step 4 indicate the final clusters detected using our network clustering method. }
    \label{illu:workflow}
\end{figure}

Our method integrates time series prediction, network construction, community detection, and cryptocurrency selection. Figure \ref{illu:workflow} outlines the sequential steps and components of our framework for portfolio construction.

\subsection{Cryptocurrency price return}
\label{sec:CPR}

We denote $P_{i,t}$ as the price (in USD) of cryptocurrency $i$ at time $t$. To remove trends from the time series, we transform the original time series of prices to a time series of log returns $r_{i,t}$:
\begin{equation} 
r_{i,t} = \ln \frac{P_{i,t}}{P_{i,t-1}}
\end{equation}

The study period is composed of a training and a test period. The training period is set to $[t_0-\Delta_{T}, t_0]$, where $\Delta_{T} = 351$ days. The performance of the portfolios is evaluated over an investment horizon spanning $[t_0,t_0 + \Delta_{h}]$, with $\Delta_{h}$ varying from 1 to 14 days (Fig.\ref{illu:workflow}, step 1). We adopt a rolling scheme in which $t_0$ advances by 30 calendar days for each subsequent study period.

\subsection{Cryptocurrency price prediction}
\label{sec:CPP}

We conducted a comparative analysis of price predictive models, including Auto-Regressive Integrated Moving Average (ARIMA), Long Short-Term Memory (LSTM), and a Naïve method for each cryptocurrency. The Auto-Regressive Integrated Moving Average (ARIMA) model is a generalised model of Auto-Regressive Moving Average (ARMA) that combines Auto-Regressive (AR) and Moving Average (MA) components to build a composite model for return forecasting.

An ARIMA($p,d,q$) model is defined such that $p$ is the order of the auto-regressive model, $d$ is the order of differencing, and $q$ is the order of the moving average model. The fitted parameters are estimated using the minimum Akaike information criterion (AIC)~\cite{akaike1974new}.

The Long Short-Term Memory (LSTM) is a Recurrent Neural Network (RNN) designed to retain information from previous time steps for future predictions. We fit the historical data using the TensorFlow LSTM model \cite{abadi2016tensorflow} and tune the essential hyper-parameters, using the Mean Squared Error (MSE) loss function. These include the number of units, defining the dimension of hidden states in the LSTM layer; the dropout rate, which determines the probability of training each node in a layer; the learning rate, controlling the adaptation speed of the model adjusted, and the sequence length, representing the number of time steps input into the LSTM network. 
MSE remains the dominant loss for LSTM-based price forecasting because it is
smooth, convex in the model output, and its gradient is proportional to the
prediction error, which facilitates stable back-propagation. 
Deep-learning studies on financial time series use MSE both as the training objective and as our out-of-sample accuracy measure, while hyper-parameter tuning can be further refined by tracking alternative metrics such as mean absolute (percentage) error or directional accuracy to test robustness under evaluation criteria beyond MSE~\cite{sezer2020financial}.

The Na\"ive method serves as a reference model in our study, employing a prediction method where the last observed value within the training period $[t_0-\Delta_{T}, t_0]$ is used for the prediction.
It assumes that the behaviour of the cryptocurrency log-returns remains unchanged over the forecast horizon, that is, the predicted log return for a future time $t_h$ ($\hat{r}_{i,t_h}$) is set equal to the log return at the last time in the training period $t_0$, i.e., $\hat{r}_{i,t_h}$ = $r_{i,t_0}$.

The Mean Squared Error (MSE) of each cryptocurrency $i$ at $t_h$ day is used to evaluate the performance of the prediction model. 
\begin{equation}  
MSE_{i,t_h} = (r_{i,t_h} - \hat{r}_{i,t_h})^{2}
\end{equation}

\subsection{Cryptocurrency correlation network}
\label{sec:CN}

A systematic approach is used to evaluate the price relationships between different cryptocurrencies (Fig.\ref{illu:workflow}, step 2). We used a fixed period of $\Delta_c = 30$ days to calculate correlations between pairs of cryptocurrencies, which allows us to track the dynamic interactions between different cryptocurrencies over time \cite{rocha2017sampling}. 

We use the Pearson correlation coefficient $\rho_{ij}$ between log returns of cryptocurrencies $i$ and $j$ over the fixed period $\Delta_c$ to estimate the interdependency between them:
\begin{equation}
\rho_{ij} = \frac{\sum_t \left( r_{i,t} - \bar{r}_i \right) \left( r_{j,t} - \bar{r}_j \right)}{\sqrt{\sum_t \left( r_{i,t} - \bar{r}_i \right)^2 \sum_t \left( r_{j,t} - \bar{r}_j \right)^2}}
\end{equation}
where $\bar{r}_i$ and $\bar{r}_j$ are the mean log returns of cryptocurrencies $i$ and $j$ respectively. We then transform correlations to distances via a standard nonlinear procedure~\cite{mantegna1999hierarchical}:
\begin{equation}
d_{ij} = \sqrt{2 \left( 1 - \rho_{ij} \right)}
\end{equation}
which maps fully correlated pairs ($\rho_{ij} = 1$) to zero distance and fully anti-correlated pairs ($\rho_{ij} = -1$) to the maximum distance. This transformation is strictly decreasing in $\rho_{ij}$. We finally construct a weighted network with nodes $i = \left\{1,2,3,..., N \right\}$ representing cryptocurrencies, and edges representing the distance $d_{ij}$ between cryptocurrencies.

\subsection{Network clustering}
\label{sec:NCP}

Modular networks consist of groups of nodes, called network communities, in which nodes in the same community are more densely connected than those in different communities.
Modularity ($Q$) quantifies the extent to which a network can be divided into these distinct communities (eq.\ref{eq:Q}). It measures the number of edges inside communities relative to what would be expected at random.

\begin{equation}
Q = \frac{1}{2m} \sum_{i,j} \left[ \rho_{ij} - \frac{k_i k_j}{2m} \right] \delta(c_i, c_j)
\label{eq:Q}
\end{equation}
where $\rho_{ij}$ represents the weight of the edge between nodes $i$ and $j$. The weighted degree $k_i$ and $k_j$ are the sum of the weights of the edges connected to nodes $i$ and $j$, respectively. The total sum of all edge weights in the network is denoted by $m$. The function $\delta(c_i, c_j)$ equals 1 if nodes $i$ and $j$ belong to the same community, and 0 otherwise, where $c_i$ and $c_j$ denote the community assignments for nodes $i$ and $j$, respectively. 
We apply the Louvain algorithm to identify the communities in the correlation network defined in each time interval $\Delta_{c}$~\cite{blondel2008fast}.

The Louvain algorithm is stochastic, and thus, we designed a method for detecting temporally stable clusters of highly correlated cryptocurrencies. The community detection procedure was repeated $n=30$ times, where each time, the detecting interval $\Delta_{c}$ was shifted backwards by one day to obtain $n$ networks (Fig.\ref{illu:workflow}, step 2). When predicted information is incorporated, $n$ networks spanning $[t_0+\Delta_{h}-\Delta_{c}-n+1, t_0+\Delta_{h}-n+1], [t_0+\Delta_{h}-\Delta_{c}-n+2, t_0+\Delta_{h}-n+2], \dots,[t_0+\Delta_{h}-\Delta_{c}, t_0+\Delta_{h}]$. Otherwise, they span $[t_0-\Delta_{c}-n+1, t_0-n+1], [t_0-\Delta_{c}-n+2, t_0-n+2], \dots,[t_0-\Delta_{c}, t_0]$. We aggregated community detection outcomes across $n$ time intervals to build an $N\times N$ similarity matrix $S$. All entries $s_{ij}$ of $S$ start with 0 and are incremented by 1 each time cryptocurrencies $i$ and $j$ are identified within the same community during a given detection interval $\Delta_{c}$. The normalised similarity score $\tilde{s}_{ij}$ between any pair of cryptocurrencies is calculated by $\tilde{s}_{ij} = \frac{s_{ij}}{n}$, where $s_{ij}$ is an element of the aggregated matrix, and $n$ is the total number of time intervals. $\tilde{s}_{ij}$ ranges from 0 to 1, where 0 indicates that the pair of cryptocurrencies $i$ and $j$ never appear in the same community, and 1 indicates consistent co-membership across all periods. This normalised matrix $\tilde{S}$ serves as the basis for constructing a similarity network $\Gamma$ (Fig.\ref{illu:workflow}, step 3), where nodes still represent cryptocurrencies but now edges reflect the frequency of co-occurrence in the same communities. 

The final clustering structure (consensus) is obtained through a hierarchical decomposition of the matrix $\tilde{S}$. We apply a majority threshold $\theta_M=0.5$ to the normalised network $\tilde{S}$ to ensure that two cryptocurrencies are considered part of the same stable cluster only if they have appeared at least 50\% of the times together, removing all edges with $\tilde{s}_{ij} < \theta_M$. The resulting connected components in network $\Gamma$ are isolated groups (or clusters) of nodes representing a stable cluster of interdependent cryptocurrencies (see clusters of step 4 in Fig.\ref{illu:workflow}). This filtering step averages the detected communities over different realisations of the stochastic algorithm and shifts training periods, thus removing spurious memberships and focusing on strong inter-dependencies.

\subsection{Portfolio forming strategy}
\label{sec:PFS}

The stable clustered structure is used to identify groups of strongly interdependent cryptocurrencies. To ensure that the portfolio incorporates a diverse set of cryptocurrencies, each representing a different aspect of market behaviour and potential investment value, we select one cryptocurrency uniformly at random from each cluster to be included in the portfolio (Fig.\ref{illu:workflow}, step 4).

\begin{figure}[h]
\centering
    \includegraphics[scale=1.0]{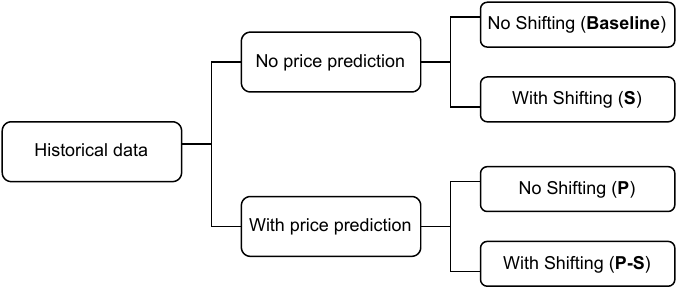}
    \caption{Flowchart illustrating the strategies employed for detecting stable clusters of highly correlated cryptocurrencies. The strategies are divided based on whether they utilise predicted data and whether they include shifting in the training periods.}
    \label{fig:flowchart}
\end{figure}

We propose four strategies that utilise both historical and predicted returns to identify stable clusters within the price correlation network (Fig.\ref{fig:flowchart}).
The baseline (Baseline) strategy employs purely historical price data to identify community structures without changing the detecting interval $\Delta_{c}$. It repeatedly applies community detection within the same interval across $n$ different initial conditions to identify stable clusters. The (S) strategy also relies solely on historical price data but incorporates a shifting mechanism where the detecting interval $\Delta_{c}$ is systematically shifted backwards by one day for each realisation of the community detection algorithm, over $n$ intervals. This mechanism exploits historical patterns reflected on average over varied periods. The (P) strategy employs prediction models to forecast future individual returns and incorporate them into the network construction process. Community detection is performed within the same interval $\Delta_{c}$, including the test period and repeated $n$ times, analogously to the baseline.
The (P-S) strategy uses predicted prices obtained from the prediction model and includes the shifting for clustering detection. This method leverages both return predictions and historical movements to identify persistent correlations between cryptocurrencies and, consequently, more stable, intrinsic inter-dependencies between them.

\subsection{Cryptocurrency portfolio}
\label{sec:CP}

Modern Portfolio Theory (MPT) provides a robust framework for distributing asset weights by evaluating the variance and correlations among assets. This framework is effective not only for traditional but also for non-traditional asset classes, such as cryptocurrencies. 
To navigate the trade-off between risk and return, our study incorporates the Sharpe Ratio (SR) in the portfolio optimisation (eq.\ref{eq:SR}). The Sharpe Ratio measures the excess return per unit of risk relative to a risk-free rate ($r_f$). 

\begin{equation} 
\text{SR} = \frac{\mathbb{E}\left[r_{\text{portfolio}} - r_f\right]}{\sqrt{\text{Var}\left[r_{\text{portfolio}} - r_f\right]}} = \frac{\mathbb{E}\left[r_{\text{portfolio}}\right] - r_f}{\sigma_{\text{portfolio}}} 
\label{eq:SR}
\end{equation}
where $\mathbb{E}\left[r_{\text{portfolio}}\right]$ is the expected portfolio return, and $\sigma_{\text{portfolio}}$ is the standard deviation of the portfolio return, representing the total risk of the portfolio. The risk-free rate is here set to $r_f=0.02$, which represents the average annualised return of 10-year U.S. Treasury bonds (standard benchmark for a risk-free investment). 

The expected return of the portfolio, $\mathbb{E}\left[r_{\text{portfolio}}\right]$, is calculated by taking the weighted average of the expected returns of individual assets in the portfolio, where $w_i$ represents the weight assigned to each asset $i$ and $\mu_i$ represents the expected return of asset $i$:
\begin{equation} 
\mathbb{E}\left[r_{\text{portfolio}}\right] = \sum_{}^{} w_i \mu_i 
\end{equation}

The portfolio’s volatility $\sigma_{\text{portfolio}}$ is computed using the covariance matrix of the cryptocurrencies' returns $\Sigma$ and the vectors of weights for each cryptocurrency $\mathbf{w}$, with $\mathbf{w} = (w_{1},...,w_{i}...,w_{n})$, $w_i \in [0,1]$:
\begin{equation} 
\sigma_{\text{portfolio}} = \sqrt{\mathbf{w}^{T} \Sigma \mathbf{w}} 
\end{equation}

By integrating the SR maximisation approach within the framework of MPT, our method not only targets higher returns but also ensures that these returns are achieved with a calibrated risk level, suitable for the volatile nature of cryptocurrency markets. The optimisation of asset weights in the portfolio, given a selected set of cryptocurrencies, is:
\begin{equation} 
\max_{\mathbf{w}} \frac{\bm{\mu}^{T}\mathbf{w} - r_f}{\sqrt{\mathbf{w}^{T}\Sigma\mathbf{w}}} 
\label{eq:sharpe_opt}
\end{equation} 
subject to $\mathbf{e}^{T}\mathbf{w} = 1$, where $\bm{\mu}= (\mu_1, \mu_2, \dots)$ is the vector of expected returns for each cryptocurrency and $\mathbf{e} = (1, 1, \dots)$ denotes the unity vector.

Because our estimation window is \(\Delta_c=30\) days and the asset universe can be comparably large, the sample covariance matrix \(C\) is typically poorly conditioned. Directly inverting such a matrix to compute optimal portfolio weights leads to unstable solutions and extreme leverage~\cite{pantaleo2011improved}. To regularise the estimate, we adopt the Ledoit–Wolf shrinkage estimator~\cite{ledoit2004well},
\begin{equation}
  \hat{\Sigma}_{\mathrm{LW}}
  =\delta^\star F+(1-\delta^\star)C,\qquad 0\le\delta^\star\le1
\end{equation}
where $F=\bar{\sigma}^2 I$ is the constant-variance target, given by the identity matrix scaled by the cross-sectional average variance, and $\delta^\star$ is the data-driven shrinkage intensity derived to minimise the mean-squared error $\mathbb{E}\lVert\hat{\Sigma}_{\mathrm{LW}}-\Sigma\rVert_F^2$, guaranteeing positive-definiteness and computational efficiency.

The PyPortfolioOpt library~\cite{Martin2021} in Python is used for computing expected returns, shrinkage-based covariance matrices, and optimal weights under MPT. The EfficientFrontier module is used to construct the Sharpe-ratio maximising portfolio, where the process of assigning weights to various assets is streamlined, aiming to maximise risk-adjusted returns.

\subsection{Portfolio performance metrics}
\label{sec:ppm}

A set of complementary metrics is employed to evaluate the profitability, consistency, and risk-adjusted performance of portfolios constructed under the proposed strategies.

The Average Trade (AT) quantifies the mean profit or loss per trade~\cite{kaufman2013trading}: 
\begin{equation} 
AT = \frac{1}{M} \sum_{i_{tr}=1}^{M}P_{i_{tr}}
\end{equation}
where $P_{i_{tr}}$ represents the profit or loss for trade $i_{tr}$, which is equivalent to the simple return $r_{i_{tr}}$ when the initial investment is normalised to 1, and $M$ denotes the total number of trades. 
Positive values correspond to profitable trades and negative values to losses.

The Win Rate (WR) measures the ability of a trading strategy to generate profitable trades, defined as the percentage of trades with positive returns out of all trades~\cite{pardo2011evaluation}: 
\begin{equation} 
WR = \frac{M_{\text{win}}}{M}
\end{equation}
where $M_\text{win}$ is the number of winning trades, i.e., trades with a positive return. 

The Profit Factor (PF) assesses the ratio of the total profits to the total losses from all trades~\cite{kaufman2013trading,pardo2011evaluation}:
\begin{equation} 
PF = \frac{\sum_{i_{tr}=1}^{M}P_{i_{tr}}^{+}}{\sum_{i_{tr}=1}^{M}P_{i_{tr}}^{-}}
\end{equation}
where $P_{i_{tr}}^{+}$ is the profit from trade $i_{tr}$, and $P_{i_{tr}}^{-}$ is the loss from trade $i_{tr}$. $P_{i_{tr}}^{+}$ and $P_{i_{tr}}^{-}$ are non-negative numbers and a $PF>1$ implies that gross profits exceed gross losses.

To capture downside and asymmetric risk characteristics, we study three metrics.
Value-at-Risk (VaR) provides a benchmark measure of potential portfolio losses under normal market conditions~\cite{dowd2007measuring}. It quantifies the maximum expected loss over a specified holding period $\Delta_h$ at a confidence level $\alpha$, defined as the $\alpha$-quantile of the portfolio return distribution:
\begin{equation}
\text{VaR}_{\alpha}(r_p) = \inf \{ r \in \mathbb{R} : F_{r_p}(r) \geq \alpha \}
\end{equation}
where $F_{r_p}(r)$ denotes the cumulative distribution function of portfolio returns. 
We estimate the distribution $F_{r_p}(r)$ empirically from portfolio returns across all periods. $\text{VaR}_{0.05}$ denotes the threshold below which the worst $5\%$ of daily outcomes fall. Lower (more negative) values indicate higher downside risk.

The Marginal Expected Shortfall (MES) captures the expected portfolio loss conditional on severe market downturns, providing a tail-risk sensitivity measure~\cite{acharya2017measuring}. It is defined as the expected return of the portfolio when the market experiences losses beyond its VaR threshold:
\begin{equation}
MES = \mathbb{E}\left[ r_p \mid r_m \leq \text{VaR}_{\alpha}(r_m) \right]
\end{equation}
where $r_p$ and $r_m$ denote portfolio and market returns, respectively. More negative MES values signal higher sensitivity to systemic risk. The market return $r_m$ is defined as the equal-weighted portfolio of all cryptocurrencies $N$, providing a proxy for aggregate market behaviour and mitigating market-cap concentration effects.

The Omega ratio provides a risk–reward measure that accounts for all moments of the return distribution~\cite{keating2002introduction}. It is the ratio of the probability-weighted gains to the probability-weighted losses relative to a chosen threshold return $r_t$:
\begin{equation}
\Omega(r_t) = 
\frac{\int_{r_t}^{\infty} [1 - F_{r_p}(r)] \, dr}
{\int_{-\infty}^{r_t} F_{r_p}(r) \, dr}
\end{equation}
where $F_{r_p}(r)$ denotes the cumulative return distribution and $r_t = 0$ is the benchmark threshold. $\Omega(r_t)$ measures the expected gain per unit of expected loss beyond the reference level $r_t$. $\Omega(r_t)>1$ indicates favourable risk–reward asymmetry, i.e. more probability-weighted mass above the threshold than below.

\subsection{Cryptocurrency Price Data}
\label{data}
The dataset comprises daily closing market prices (at midnight) of 5,450 cryptocurrencies collected from various online sources. We extracted the top 1,000 currencies with the highest market cap on 22.02.2022 and then retained those persistently active (being traded) from 15.11.2017 to 15.04.2022 ($T_{\text{total}}=1,613$ days). Cryptocurrencies with missing prices for more than ten days are excluded. The final data set contains a time series of prices (in USD) for $N=157$ cryptocurrencies. The full list of cryptocurrencies used in our study is given in the supplementary information.

\section{Results and Discussion}

In this section, we first present descriptive statistics of cryptocurrency price returns, then evaluate the performance of the prediction models, analyse the community structure of the correlation network and the properties of the detected clusters, and finally assess the performance of the proposed portfolios.

\subsection{Descriptive statistics}
\label{DS}

The descriptive statistics of the log return for a representative subset of assets provide an overview of key distributional properties such as volatility, skewness, and kurtosis, which are relevant for price forecasting and network analysis.

\begin{table}[h]
\sisetup{table-format=-1.4} 
    \centering
    \small
    \begin{tabular}{lrrrrrrrrrr}
    {} &        \textbf{BTC} &        \textbf{ETH} &        \textbf{LTC} &       \textbf{USDT} &        \textbf{XRP} &       \textbf{OCN} &        \textbf{DLT} &        \textbf{ENG} &        \textbf{ETP} &       \textbf{FUEL} \\
    \midrule
    \textbf{T}                   &  1612 &  1612 &  1612 &  1612 &  1612 &  1612 &  1612 &  1612 &  1612 &  1612 \\
    \textbf{Mean ($\times 10^{-3}$)}                &     1.065 & 1.378 & 0.349 & -0.003 & 0.823 & 0.770 & -2.732 & -1.800 & -2.110 & -2.708 \\ 
    \textbf{Std deviation}  &     0.042 &     0.053 &     0.058 &     0.004 &     0.067 &     0.176 &     0.094 &     0.090 &     0.078 &     0.375 \\
    \textbf{Minimum}             &    -0.480 &    -0.570 &    -0.466 &    -0.057 &    -0.541 &    -0.581 &    -0.960 &    -0.769 &    -0.829 &    -2.943 \\
    \textbf{Median ($\times 10^{-3}$)}              &     1.691 & 1.752 & -0.254 & 0.000 & 0.178 & -4.167 & -1.321 & -2.219 & -3.151 & -3.774 \\ 
    \textbf{Maximum}             &     0.203 &     0.246 &     0.426 &     0.053 &     0.618 &     6.099 &     0.809 &     0.889 &     0.591 &     2.920 \\
    \textbf{Skewness}            &    -0.922 &    -0.927 &    -0.059 &     0.066 &     0.917 &    25.969 &    -0.374 &     0.364 &    -0.417 &    -0.070 \\
    \textbf{Kurtosis}            &    12.743 &    10.496 &     9.223 &    75.109 &    15.453 &   898.229 &    15.624 &    13.972 &    15.725 &    32.004 \\
    \textbf{ADF test statistic} &   -28.080 &   -27.494 &   -15.159 &   -10.320 &   -27.452 &   -42.422 &   -11.396 &   -43.292 &   -27.504 &    -7.631 \\
    \textbf{ADF test p-value} & \textless 0.001 & \textless 0.001 & \textless 0.001 & \textless 0.001 & \textless 0.001 & \textless 0.001 & \textless 0.001 & \textless 0.001 & \textless 0.001 & \textless 0.001 \\
    \bottomrule
    \end{tabular}
    \caption{Descriptive statistics and augmented Dickey–Fuller (ADF) test without trend of log returns ($r_{i,t}$) taken at $T$ times (days) for the top 5 and bottom 5 cryptocurrencies according to their average market capitalisation over the whole study period. The p-values of the ADF test reject the null hypothesis, which indicates that the time series is stationary. Time $T$ is given in days, and the Mean, Std deviation, Minimum, Median, and Maximum of the log returns are calculated in USD.}
    \label{tab:descriptive}
\end{table}

Table \ref{tab:descriptive} reports the log return $r_{i,t}$ for the top 5 (BTC: Bitcoin; ETH: Ethereum; LTC: Litecoin; USDT: Tether; XRP: Ripple) and bottom 5 (OCN: Odyssey; DLT: Agrello; ENG: Enigma; ETP: Metaverse; FUEL: Etherparty) cryptocurrencies according to their average market capitalisation during the study period.
Bitcoin (BTC) and Ethereum (ETH) show positive mean returns, whereas lower market-cap cryptocurrencies such as DLT and FUEL exhibit negative mean returns. The standard deviation indicates higher volatility for lower market cap cryptocurrencies, with FUEL having the highest standard deviation. 
The skewness and kurtosis indicate non-normality, with heavy tails and frequent extreme values, particularly for cryptocurrencies such as OCN.
Augmented Dickey-Fuller (ADF) test results confirm that the log returns $r_{i,t}$ of all examined cryptocurrencies are stationary, rejecting the null hypothesis at $p<.001$.

\begin{table}[h]
\sisetup{table-format=-1.4} 
    \centering
    \small
    \begin{tabular}{l|rrrrrrrrrr}
    {} &     \textbf{BTC} &    \textbf{ETH} &     \textbf{LTC} &     \textbf{USDT} &      \textbf{XRP} &      \textbf{OCN} &      \textbf{DLT} &      \textbf{ENG} &      \textbf{ETP} &    \textbf{FUEL} \\
    \midrule
    \textbf{BTC}  &  \textbf{1} &  \textbf{0.57} &  \textbf{0.77} &   0.02 &   \textbf{0.56} &   0.18 &   0.30 &   0.45 &   \textbf{0.53} &  0.12 \\
    \textbf{ETH} &  \textbf{0.57} &  \textbf{1} &  \textbf{0.58} &   0.03 &   0.49 &   0.13 &   0.37 &   0.38 &   0.42 &  0.12 \\
    \textbf{LTC}  &  \textbf{0.77} &  \textbf{0.58} &  \textbf{1} &   0.06 &   \textbf{0.64} &   0.19 &   0.32 &   0.44 &   \textbf{0.52} &  0.11 \\
    \textbf{USDT} &  0.02 &  0.03 &  0.06 &   \textbf{1} &  -0.02 &  -0.06 &  -0.03 &  -0.02 &  -0.05 &  0.00 \\
    \textbf{XRP}  &  \textbf{0.56} &  0.49 &  \textbf{0.64} &  -0.02 &   \textbf{1} &   0.13 &   0.27 &   0.35 &   0.44 &  0.12 \\
    \textbf{OCN}  &  0.18 &  0.13 &  0.19 &  -0.06 &   0.13 &   \textbf{1} &   0.08 &   0.11 &   0.10 &  0.03 \\
    \textbf{DLT}  &  0.30 &  0.37 &  0.32 &  -0.03 &   0.27 &   0.08 &   \textbf{1} &   0.26 &   0.32 &  0.07 \\
    \textbf{ENG}  &  0.45 &  0.38 &  0.44 &  -0.02 &   0.35 &   0.11 &   0.26 &   \textbf{1} &   0.36 &  0.08 \\
    \textbf{ETP}  &  \textbf{0.53} &  0.42 &  \textbf{0.52} &  -0.05 &   0.44 &   0.10 &   0.32 &   0.36 &   \textbf{1} &  0.08 \\
    \textbf{FUEL} &  0.12 &  0.12 &  0.11 &   0.00 &   0.12 &   0.03 &   0.08 &   0.08 &   0.08 &  \textbf{1} \\
    \bottomrule
    \end{tabular}
    \caption{Correlation matrix ($\rho_{ij}$) of log returns for the top 5 and bottom 5 cryptocurrencies ranked according to their average market capitalisation over the whole study period. Correlations larger than 0.5 are in bold.}
    \label{tab:correlation}
\end{table}

Table \ref{tab:correlation} shows the price correlation matrix between pairs of cryptocurrencies. The highest correlations are observed between the high-cap cryptocurrencies such as BTC, ETH, and LTC, indicating that these cryptocurrencies often move together and reflect common market factors that can be exploited by predictive models.
Conversely, Tether (USDT) shows relatively low correlations with other cryptocurrencies, reflecting its function as a stablecoin pegged to the U.S. dollar. As traders move into stablecoins like USDT during periods of stress, correlation patterns across the broader market can shift abruptly, and these sudden changes in correlation can pose additional challenges for predictive models, which may struggle to capture the rapidly evolving dynamics of the market.

\subsection{Performance price prediction}
\label{PPP}

We evaluated the performance of three log-return prediction methods (see Sec.\ref{sec:CPP} for details) across $n_s = 42$ study periods, each containing a training period $\Delta_T = 351$ days and a test horizon of up to 14 days. 

\begin{figure}[h]
\centering
    \includegraphics[scale=0.45]{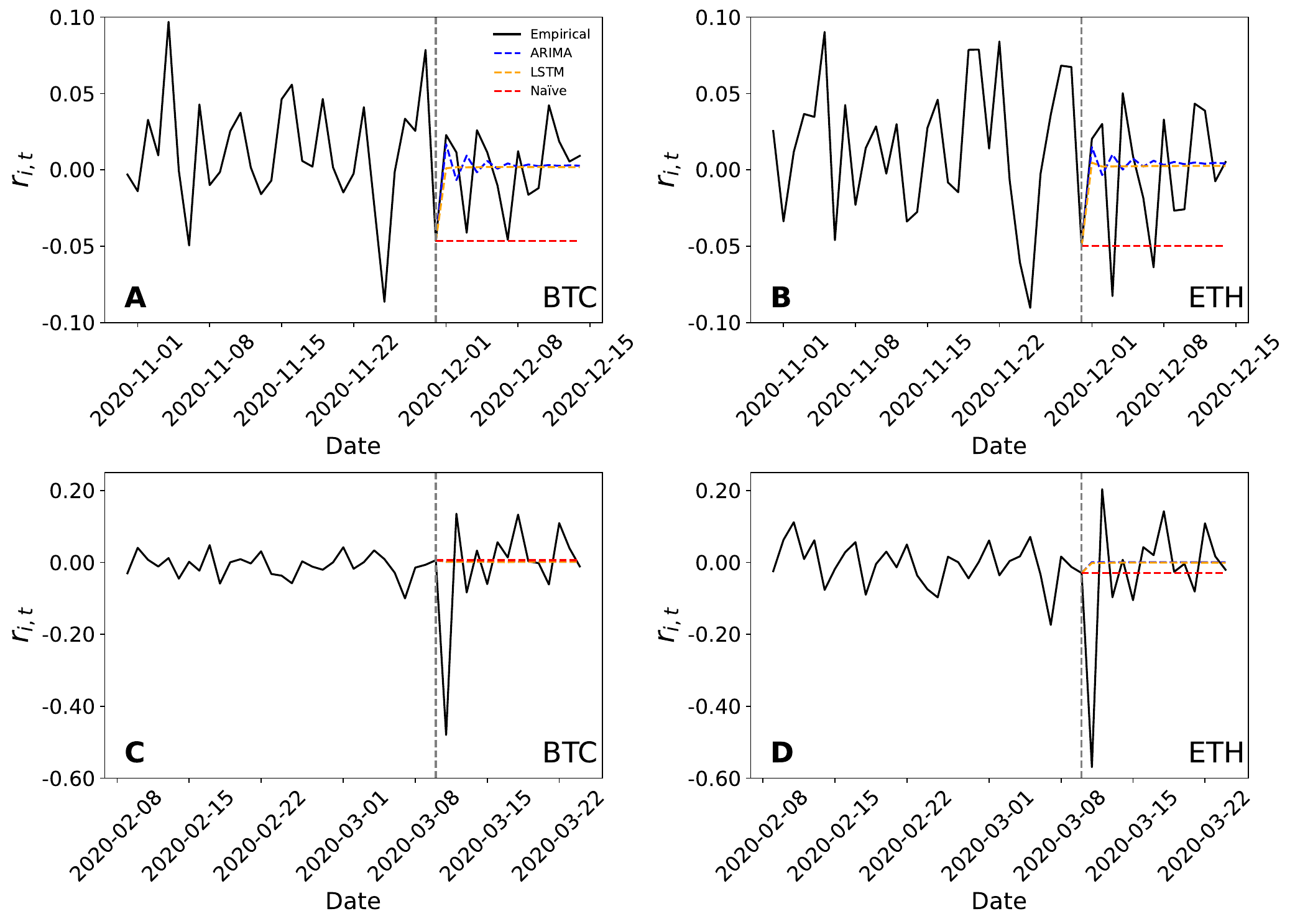}
    \caption{Time series of log returns for two cryptocurrencies (A) BTC and (B) ETH for each model of predicting horizon from 1 to 14 days, using price data during a period without market disruptions, and (C) BTC and (D) ETH during an external shock (COVID-19 outbreak). The vertical black dashed line shows time $t_0$, which separates the training and prediction periods. The dashed lines represent price predictions according to different models. }
    \label{fig:prediction}
\end{figure}

Figure~\ref{fig:prediction} shows the log return predictions for BTC and ETH over horizons from 1 to 14 days, for periods with and without external shocks. In periods of relative stability, day-by-day MSE values remain relatively low for both cryptocurrencies. For BTC (Fig.~\ref{fig:prediction}A), the mean MSE over the 1–14 day horizon is \(5.80\times 10^{-4}\) for LSTM, \(6.34\times 10^{-4}\) for ARIMA, and \(2.98\times 10^{-3}\) for the Na\"ive method. For ETH (Fig.~\ref{fig:prediction}B), the corresponding mean MSEs are \(1.50\times 10^{-3}\) (LSTM), \(1.65\times 10^{-3}\) (ARIMA), and \(4.02\times 10^{-3}\) (Na\"ive), confirming that both ARIMA and LSTM improve on the simple benchmark in calm market conditions.

The performance changes during turbulent periods, such as around the World Health Organisation’s declaration of COVID-19 as a global pandemic on 11 March 2020, when uncertainty and volatility surged in cryptocurrency markets. Average MSEs over the prediction horizon rose substantially for both assets. For BTC (Fig.~\ref{fig:prediction}C), ARIMA and LSTM both reach mean MSEs around \(2.14\times 10^{-2}\), with the Na\"ive method performing comparably at \(2.16\times 10^{-2}\). Similarly, for ETH (Fig.~\ref{fig:prediction}D), mean MSEs are \(3.07\times 10^{-2}\) (ARIMA), \(3.06\times 10^{-2}\) (LSTM), and \(3.00\times 10^{-2}\) (Na\"ive), indicating that all models struggle under large volatility.

Overall, ARIMA and LSTM outperform the Naïve method in stable environments, but this advantage largely disappears during high-volatility episodes, where the Naïve benchmark becomes comparatively competitive. The pronounced increase in MSE during turbulent periods highlights the difficulty of forecasting log returns under sudden market shifts and underscores the need for predictive strategies that remain robust across both normal and stress regimes.

\begin{table}[h]
\sisetup{table-format=1.2}   
\centering
\small
   \begin{tabular}{l|rrrrrrrrrrrrrrr}
        \textbf{$\Delta_h$} (days) & \textbf{1} & \textbf{2} & \textbf{3} & \textbf{4} & \textbf{5} & \textbf{6} & \textbf{7} & \textbf{8} & \textbf{9} & \textbf{10} & \textbf{11} & \textbf{12} & \textbf{13} & \textbf{14} \\ \hline
        \textbf{LSTM} & 1.58 & 1.96 & 1.97 & 1.70 & 1.94 & \textbf{2.00} & 1.94 & 1.30 & 2.27 & 2.04 & 8.72 & \textbf{2.11} & 2.27 & 1.57 \\ 
        \textbf{ARIMA} & \textbf{1.35} & \textbf{1.92} & \textbf{1.84} & \textbf{1.54} & \textbf{1.71} & 2.09 & \textbf{1.90} & \textbf{1.29} & \textbf{2.25} & \textbf{1.94} & \textbf{8.60} & 2.13 & \textbf{2.15} & \textbf{1.48} \\ 
        \textbf{Naïve} & 2.88 & 3.21 & 3.06 & 2.96 & 3.26 & 3.78 & 2.80 & 2.43 & 4.52 & 3.91 & 10.10 & 3.76 & 3.68 & 2.97 \\ \hline
    \end{tabular}
    \caption{The average of the median MSE ($\e{-3}$) for prediction horizons $\Delta_h = 1, 2 ,... 14$ days. This metric is obtained by first taking the median MSE across all $N=157$ cryptocurrencies, followed by averaging these medians across all $n_s = 42$ study periods. The best prediction model with the smallest MSE for each day prediction is labelled in bold.}
    \label{tab:MSE_tables}
\end{table}

We evaluated the prediction accuracy of the three methods across all study periods by calculating, for each cryptocurrency and prediction horizon \(\Delta_h\), the Mean Squared Error (MSE) and then taking the median over \(\Delta_h\). The median was chosen because it is less sensitive to extreme values. These median MSEs were then averaged over all \(n_s = 42\) study periods for each method. ARIMA generally outperforms LSTM and the Na\"ive method, achieving the lowest MSE across nearly all prediction horizons (Table~\ref{tab:MSE_tables}).

\begin{figure}[h]
\centering
    \includegraphics[width=0.95\linewidth]{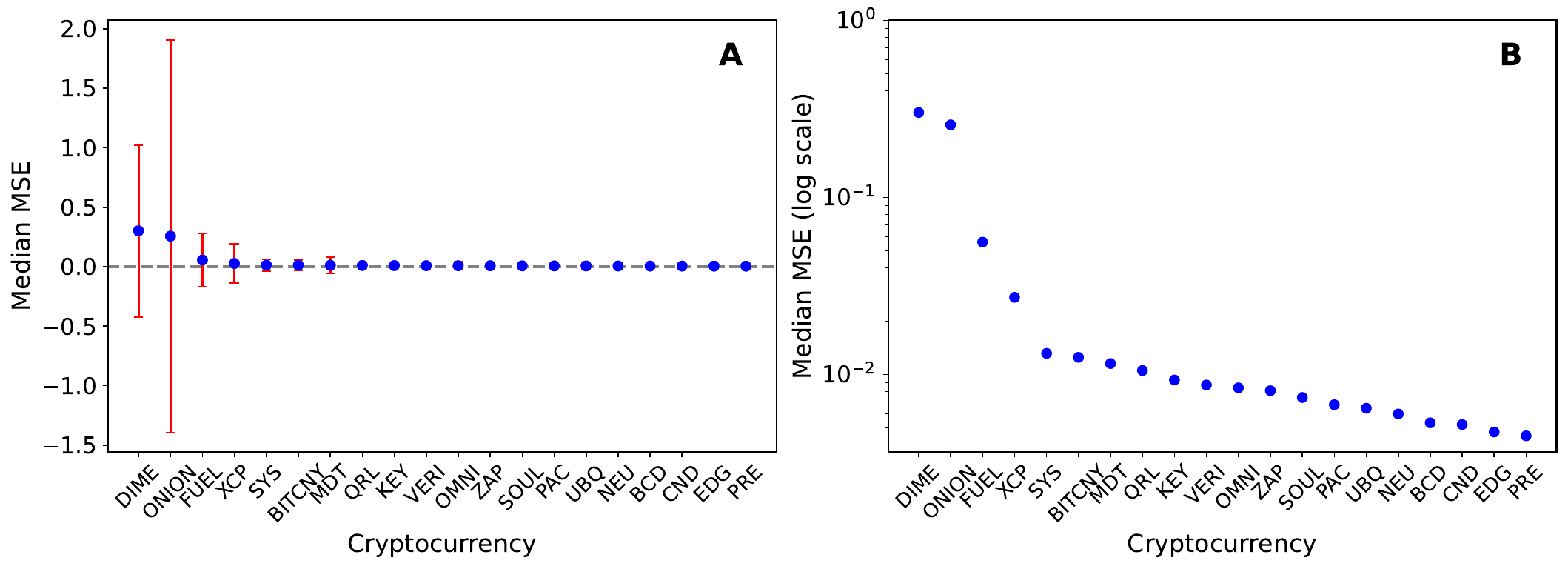}
    \caption{The top 20 cryptocurrencies with the largest average of the median MSE from the ARIMA method over all $n_s = 42$ study periods, (A) on a linear y-axis and its standard deviation, and (B) on a logarithmic y-axis.}
    \label{fig:wt_MSE_20}
\end{figure}

Figure~\ref{fig:wt_MSE_20} shows the top 20 cryptocurrencies with the largest average median MSE under the ARIMA model. Since the median operator down-weights occasional large deviations and MSE is scale-dependent, the log scale in Fig.~\ref{fig:wt_MSE_20}B clarifies the spacing among smaller error values. The results indicate that the vast majority (98.73\%) of cryptocurrencies have an average median MSE below 0.056, while DIME and ONION exhibit higher errors, signalling lower predictability. This suggests that their price dynamics may be driven by pronounced volatility, low liquidity, or idiosyncratic shocks, and may require tailored predictive strategies or cautious inclusion in model-based applications.

\subsection{Cryptocurrency cluster identification}
\label{CCI}

\begin{table}[h]
\sisetup{table-format=1.4}   
\centering
    \small
    \begin{tabular}{p{1.0cm}p{1.8cm}p{1.8cm}p{1.8cm}p{1.8cm}p{1.8cm}p{1.8cm}}
        {} & \textbf{Baseline} & \textbf{P(ARIMA)} & \textbf{S} & \textbf{P(ARIMA)-S}  \\ \hline
        \textbf{$\left\langle  Q \right\rangle$} & 0.124 & 0.123    & 0.121 & 0.122  \\ 
        \textbf{$\sigma$} & 0.087 & 0.084 & 0.071 & 0.069
  \\ \hline
    \end{tabular}
    \caption{Average modularity ($\left\langle Q \right\rangle$) and the corresponding standard deviation ($\sigma$) over all $n_s = 42$ study periods. This metric is obtained by first computing the mean $Q$ across the $n = 30$ realisations for each study period, and then averaging those means over all $n_s = 42$ study periods.}
    \label{tab:AvgQ_tables}
\end{table}

We focus the analysis on highly correlated cryptocurrencies. Therefore, we only retain the links between cryptocurrencies in which their $d_{ij} < \sqrt{2(1-\theta_{\rho})} = 1$.
The average modularity \(\langle Q \rangle\) across \(n = 30\) realisations per period is relatively low, around 0.12 (Tab.~\ref{tab:AvgQ_tables}), which justifies using the consensus methodology that we proposed.
Among all strategies, those without sliding windows (Baseline and P(ARIMA)) yield slightly higher \(\langle Q \rangle\), whereas sliding-window strategies (S and P(ARIMA)-S) exhibit lower \(\langle Q \rangle\) but also lower standard deviations \(\sigma\), reflecting more consistent cluster definitions over time.
Low global modularity is a well-documented property of financial correlation networks, where dense connectivity and strong common factors compress community separation, particularly during periods of elevated co-movement~\cite{rakib2021structure}.
Our network consists of a single giant connected component with high overall connectivity, so \(Q\) mainly reflects the dominance of the common market mode rather than the absence of local structure. Our stable clustering procedure is specifically designed to extract these local, persistent co-movement patterns. By aggregating community membership over time and retaining only pairs that repeatedly co-occur in the same community, we recover stable clusters.
A sensitivity analysis of the modularity across a range of thresholds $d_{ij} \in \bigl[\sqrt{0.2},\, \sqrt{2}\bigr]$ (see SI) confirms that \(\theta_{\rho} = 0.5\) provides a balance between modularity strength and cluster stability.

\begin{figure}[h]
    \centering
    \includegraphics[width=\textwidth]{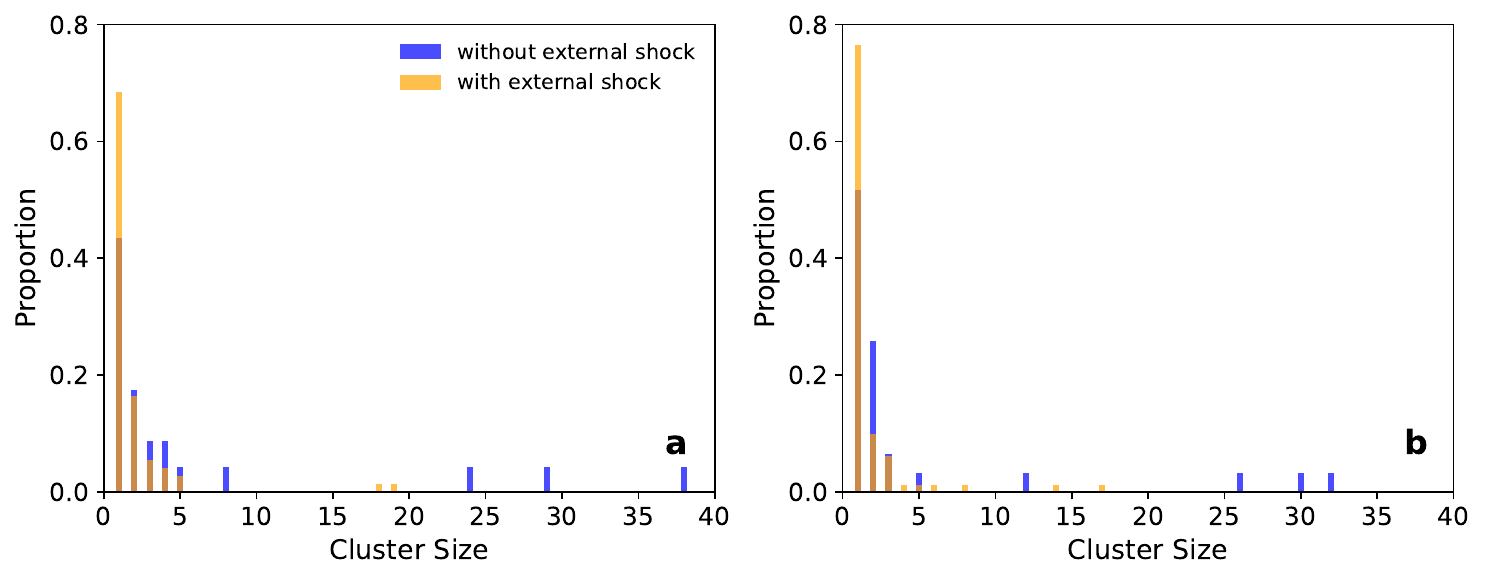}
    \caption{Size distribution of the final stable cryptocurrency clusters after applying the (A) Baseline strategy and (B) P(ARIMA) strategy, during the same study periods with and without external shocks (COVID-19 outbreak).}
    \label{fig:comdist}
\end{figure}

Figure~\ref{fig:comdist} shows the size distribution of final stable clusters constructed using the Baseline and P(ARIMA) strategies, during the same study period with and without external shocks. In both cases, the distributions are skewed toward single-node clusters, with the P(ARIMA) strategy yielding a higher number of singletons. Larger clusters, containing up to 38 nodes, also appear. The predominance of single-node clusters tends to produce more consistent cryptocurrency choices for portfolio construction, as there are fewer alternatives within each cluster and the same ones are often selected repeatedly. By contrast, larger clusters introduce greater variation in the randomised selection process, increasing portfolio diversity and influencing the resilience and variability of returns over time.

\begin{figure}[h]
    \centering
    \includegraphics[width=\textwidth]{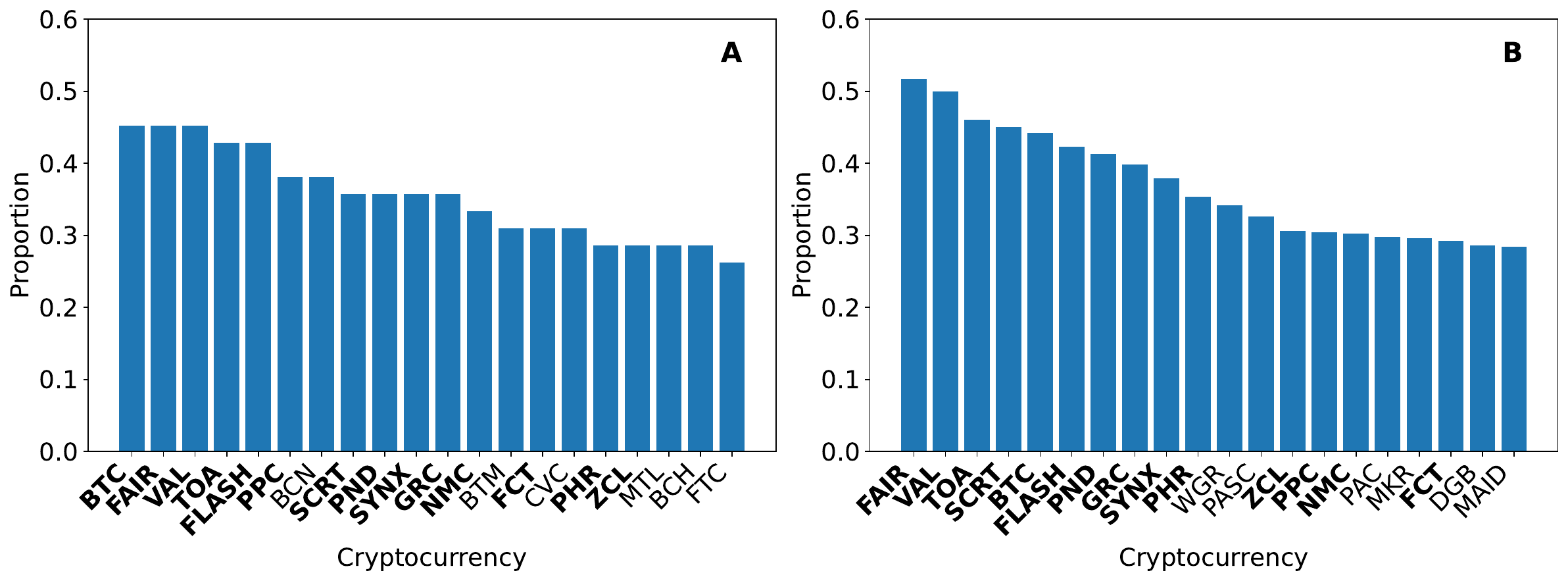}
    \caption{Top 20 cryptocurrencies ranked by the proportion of occurrence within the largest cluster identified across all $n_s = 42$ study periods for the (A) Baseline and (B) P(ARIMA) strategies. Cryptocurrencies appearing in the top 20 for both strategies are highlighted in bold.}
    \label{fig:coin_distri_largest}
\end{figure}

Figure~\ref{fig:coin_distri_largest} shows the top 20 cryptocurrencies ranked by their proportion of occurrence within the largest cluster across all \(n_s = 42\) study periods for the Baseline and P(ARIMA) strategies. Fourteen cryptocurrencies are common to the top 20 sets of both strategies. Several of these are not among the most widely recognised large-cap assets (such as BTC) yet consistently appear in the largest clusters, suggesting that they co-move strongly with BTC and other central assets. Their persistence in large clusters indicates interaction with a broad set of cryptocurrencies. By including such assets, investors gain exposure to wider segments of the market through the interdependencies captured in the clusters, helping to spread risk and tap into common underlying trends or collective market sentiment.

\subsection{Portfolio performance}
\label{PP}

\captionsetup[table]{skip=1pt} 
\begin{table}[h]
\begin{minipage}{0.46\textwidth}
\centering
\caption*{\small\textbf{Average Trade (AT) (\%)}}
\small
\begin{tabular}{c|c
>{\columncolor[HTML]{C0C0C0}}c cc}
\hline
\textbf{$\Delta_h$   (days)} & \textbf{Baseline} & \textbf{P(ARIMA)}  & \textbf{S} & \textbf{P(ARIMA)-S}  \\ \hline
\textbf{1} & \textbf{1.82} & \textbf{1.39} & \textbf{1.59} & \textbf{1.70} \\
\textbf{2} & \textbf{1.23} & \textbf{0.91} & \textbf{1.01} & \textbf{1.19} \\
\textbf{3} & \textbf{0.68} & \textbf{0.25} & \textbf{0.23} & \textbf{0.43} \\
\textbf{4} & \textbf{0.61} & \textbf{0.47} & \textbf{0.51} & \textbf{0.69} \\
\textbf{5} & \textbf{0.62} & \textbf{0.59} & \textbf{0.32} & \textbf{0.51} \\
\textbf{6} & \textbf{0.33} & \textbf{0.43} & \textbf{0.19} & \textbf{0.31} \\
\textbf{7} & \textbf{0.20} & \textbf{0.34} & \textbf{0.18} & \textbf{0.22} \\
\textbf{8} & \textbf{0.34} & \textbf{0.40} & \textbf{0.26} & \textbf{0.28} \\
\textbf{9} & \textbf{0.10} & \textbf{0.16} & \textbf{0.04} & \textbf{0.03} \\
\textbf{10} & \textbf{0.05} & \textbf{0.27} & -0.05 & -0.05 \\
\textbf{11} & \textbf{0.02} & \textbf{0.16} & -0.15 & -0.18 \\
\textbf{12} & \textbf{0.04} & \textbf{0.12} & -0.16 & -0.22 \\
\textbf{13} & \textbf{0.02} & \textbf{0.08} & -0.18 & -0.24 \\
\textbf{14} & -0.05 & \textbf{0.15} & -0.25 & -0.21 \\ 
\hline
\end{tabular}
\bigskip
\end{minipage}
\hfill
\begin{minipage}{0.46\textwidth}
\centering
\caption*{\small\textbf{Standard Deviation of AT ($\times 10^{-2}$)}}
\small
\begin{tabular}{c
>{\columncolor[HTML]{C0C0C0}}c cc}
\hline
\multicolumn{1}{c}{\textbf{Baseline}} & \multicolumn{1}{c}{\cellcolor[HTML]{C0C0C0}\textbf{P(ARIMA)}} & \multicolumn{1}{c}{\textbf{S}} & \multicolumn{1}{c}{\textbf{P(ARIMA)-S}} \\ \hline
3.88 & 3.27 & 4.18 & 3.76 \\
3.05 & 2.92 & 2.87 & 2.62 \\
2.40 & 2.19 & 2.37 & 2.13 \\
2.29 & 2.42 & 2.68 & 2.61 \\
2.02 & 1.80 & 2.02 & 1.88 \\
1.98 & 1.83 & 1.81 & 1.67 \\
1.89 & 1.63 & 1.73 & 1.72 \\
1.72 & 1.48 & 1.62 & 1.65 \\
1.56 & 1.35 & 1.44 & 1.45 \\
1.47 & 1.65 & 1.37 & 1.29 \\
1.68 & 1.37 & 1.76 & 1.74 \\
1.55 & 1.49 & 1.72 & 1.63 \\
1.54 & 1.52 & 1.71 & 1.66 \\
1.37 & 1.39 & 1.55 & 1.45 \\ 
\hline
\end{tabular}
\bigskip
\end{minipage}

\begin{minipage}{0.46\textwidth}
\centering
\caption*{\small\textbf{Win Rate (WR)}}
\small
\begin{tabular}{c|c
>{\columncolor[HTML]{C0C0C0}}c cc}
\hline
\textbf{$\Delta_h$   (days)} & \textbf{Baseline} & \textbf{P(ARIMA)} & \textbf{S} & \textbf{P(ARIMA)-S}  \\ \hline
\textbf{1} & \textbf{0.69} & \textbf{0.71} & \textbf{0.74} & \textbf{0.74}  \\  
\textbf{2} & \textbf{0.71} & \textbf{0.62}  & \textbf{0.71} & \textbf{0.69} \\  
\textbf{3} & \textbf{0.62} & \textbf{0.64}  & \textbf{0.60} & \textbf{0.64}  \\  
\textbf{4} & \textbf{0.62} & \textbf{0.67}  & \textbf{0.62} & \textbf{0.62}  \\  
\textbf{5} & \textbf{0.57} & \textbf{0.71} & \textbf{0.60} & \textbf{0.60} \\  
\textbf{6} & \textbf{0.55} & \textbf{0.64} & \textbf{0.62} & \textbf{0.67}  \\  
\textbf{7} & \textbf{0.55} & \textbf{0.64} & \textbf{0.55} & \textbf{0.55}  \\  
\textbf{8} & \textbf{0.52} & \textbf{0.62} & \textbf{0.57} & \textbf{0.57}  \\  
\textbf{9} & 0.50 & \textbf{0.52}  & 0.45 & 0.48 \\  
\textbf{10} & 0.48 & \textbf{0.55} & \textbf{0.57} & \textbf{0.57} \\  
\textbf{11} & 0.48 & \textbf{0.52} & \textbf{0.57} & 0.50 \\  
\textbf{12} & 0.45 & \textbf{0.52} & \textbf{0.55} & 0.48 \\  
\textbf{13} & 0.48 & 0.50 & \textbf{0.60} & \textbf{0.60} \\  
\textbf{14} & 0.50 & \textbf{0.60} & \textbf{0.55} & \textbf{0.57} \\ \hline
\end{tabular}
\end{minipage}
\hfill
\begin{minipage}{0.46\textwidth}
\centering
\caption*{\small\textbf{Profit Factor (PF)}}
\small
\begin{tabular}{c
>{\columncolor[HTML]{C0C0C0}}c cc}
\hline
\textbf{Baseline} & \textbf{P(ARIMA)} & \textbf{S} & \textbf{P(ARIMA)-S} \\ \hline
\textbf{4.97} & \textbf{3.35} & \textbf{3.57} & \textbf{4.48} \\
\textbf{3.28} & \textbf{2.27} & \textbf{2.77} & \textbf{3.64} \\
\textbf{2.24} & \textbf{1.36} & \textbf{1.31} & \textbf{1.70} \\
\textbf{2.09} & \textbf{1.80} & \textbf{1.69} & \textbf{2.05} \\
\textbf{2.17} & \textbf{2.47} & \textbf{1.49} & \textbf{1.96} \\
\textbf{1.51} & \textbf{1.96} & \textbf{1.32} & \textbf{1.64} \\
\textbf{1.30} & \textbf{1.70} & \textbf{1.30} & \textbf{1.37} \\
\textbf{1.61} & \textbf{1.94} & \textbf{1.49} & \textbf{1.53} \\
\textbf{1.17} & \textbf{1.32} & \textbf{1.08} & \textbf{1.06} \\
\textbf{1.09} & \textbf{1.58} & 0.91 & 0.90 \\
\textbf{1.03} & \textbf{1.39} & 0.77 & 0.73 \\
\textbf{1.06} & \textbf{1.24} & 0.77 & 0.68 \\
\textbf{1.03} & \textbf{1.15} & 0.75 & 0.66 \\
0.92 & \textbf{1.32} & 0.65 & 0.67 \\ 
\hline
\end{tabular}
\end{minipage}
\caption{Statistics of four performance indicators of the Average Trade (AT) (\%), the Standard Deviation of Average Trade (AT) ($\times 10^{-2}$), Win Rate (WR) and Profit Factor (PF) of each strategy portfolio for investment horizon holding from 1 to 14 days over all $n_s = 42$ studied periods. Performances above the break-even values for the Average Trade (0), the Win Rate (0.5), and the Profit Factor (1) are in bold. The columns of strategy P(ARIMA) that performed well across all holding periods are shaded in grey. Returns are reported on an implied daily simple scale, and for a given holding period $\Delta_h$, the corresponding $\Delta_h$-day holding-period return (cumulative return, HPR) can be obtained from the implied daily simple return $r_{\mathrm{d, \Delta_h}}$ as $HPR_{\Delta_h} = (1 + r_{\mathrm{d, \Delta_h}})^{\Delta_h} - 1$.}
\label{tab:all_indicators}
\end{table}

We evaluate the short-term performance of the portfolios built using the proposed strategies (Tab.~\ref{tab:all_indicators}). The average trade (AT) is strongly positive on day 1 across all strategies. It remains positive, though slowly decreasing, up to a 9-day holding period, after which a sharper decline is observed. By the 14th day, only the P(ARIMA) strategy still exhibits a positive AT. The standard deviation of AT tends to fall with longer holding periods, indicating that while average returns shrink, returns become more stable over time. WR also starts high, with all strategies exceeding $69\%$ probability of profitable trades on day 1. P(ARIMA) guarantees winnings (WR$\geq 0.5$) up to 14 days. PF values are consistently above 1 for all strategies in the short term, but similarly to AT and WR, decrease over longer investment horizons. Overall, P(ARIMA) shows a relatively stronger performance. The reference Baseline strategy performs well at short horizons, remaining relatively resilient even without predictive prices or sliding-window adjustments.

\begin{figure}[h]
\raggedright
    \includegraphics[scale=0.35]{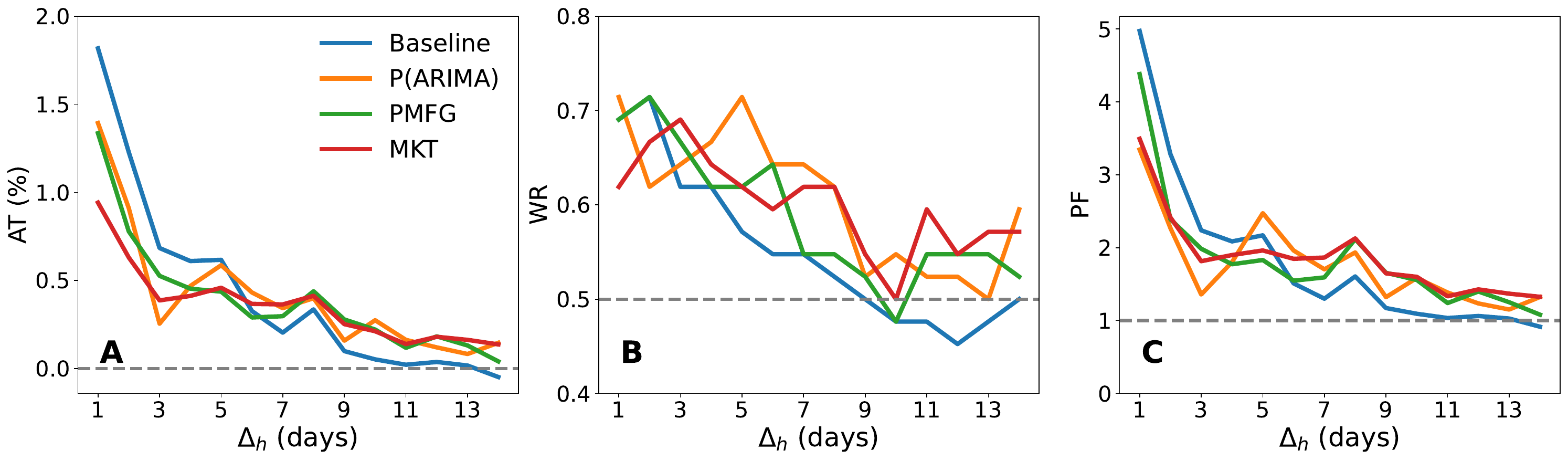}
    \caption{Performance indicators for our strategies (Baseline and P(ARIMA)) compared with two benchmark methods: a Planar Maximally Filtered Graph portfolio (PMFG), which selects assets from network-filtered correlation structures, and an equal-weight market portfolio (MKT), which assigns identical weights to all studied cryptocurrencies. Shown are (A) Average Trade (AT), (B) Win Rate (WR), and (C) Profit Factor (PF) across investment horizons ($\Delta_h$). The dashed lines indicate reference levels. Returns are measured on an implied daily simple scale.}
    \label{fig:all_indicators}
\end{figure}

Figure~\ref{fig:all_indicators} visualises the decay of performance metrics with larger $\Delta_h$ for all strategies, showing the difficulty of sustaining short-term trading edges in a highly volatile market. The equal-weight portfolio (MKT) can serve as a proxy for the broad cryptocurrency market behaviour. By observing the vertical distance between our proposed strategies and the MKT, we can isolate the excess returns generated by our approach. In fig.~\ref{fig:all_indicators}A and C, the proposed strategies, particularly the Baseline, maintain a clear gap above the market proxy up to the 5th day, indicating that the gains are not merely beta of market movement but alpha derived from structural selection. Conversely, the Planar Maximally Filtered Graph portfolio (PMFG)~\cite{tumminello2005tool,aste2005correlation} benchmark exhibits comparatively erratic performance. It fails to establish a consistent premium over the MKT, frequently converging with or dipping below the market baseline earlier than our proposed strategies.
Furthermore, the WR in fig.~\ref{fig:all_indicators}B shows that the MKT proxy is relatively competitive, and this suggests that the superior performance of our strategy is driven less by winning more often than the market, but more by achieving higher quality returns when profitable opportunities are identified.

Our portfolio strategies that incorporate time-series prediction of log returns outperform the Baseline strategy relying only on historical data, particularly over longer horizons. The P(ARIMA) strategy leverages predicted information movements to refine cluster detection based on expected market conditions, i.e. it captures evolving trends better than strategies relying only on shifting backwards historical windows. This suggests that correlation memory decreases over time. Predictive models generally demand substantial computational resources (ARIMA less than LSTM), and the incremental performance gains over the Baseline strategy may not always justify the additional cost of resources. Our findings, therefore, suggest a trade-off: while predictive and shifting strategies offer advantages in adapting to market dynamics, the simpler Baseline strategy remains attractive due to its lower computational burden with a relatively good short-term performance. More broadly, the results suggest that capturing the correlation structure and interdependencies between cryptocurrencies is as important as accurate price forecasting when designing efficient portfolios to increase returns while mitigating risk.

\begin{table}[h]
\centering
\small
\sisetup{table-format=1.2}
\begin{tabular}{
p{1.3cm}|
S[table-format=1.1]S[table-format=1.1]S[table-format=1.1]   
S[table-format=1.1]S[table-format=1.1]>{\columncolor[HTML]{C0C0C0}}S[table-format=1.1]   
S[table-format=1.1]S[table-format=1.1]S[table-format=1.1]   
S[table-format=1.1]S[table-format=1.1]S[table-format=1.1]   
}
\toprule
& 
\multicolumn{3}{c}{\textbf{Baseline}} &
\multicolumn{3}{c}{\textbf{P(ARIMA)}} &
\multicolumn{3}{c}{\textbf{S}} &
\multicolumn{3}{c}{\textbf{P(ARIMA)-S}} \\
\cmidrule(lr){2-4}\cmidrule(lr){5-7}\cmidrule(lr){8-10}\cmidrule(lr){11-13}
\textbf{$\Delta_h$ (days)} &
{\textbf{VaR$_{5\%}^{(\text{loss})}$}} & {\textbf{MES$_{5\%}^{(\text{loss})}$}} & {\textbf{$\Omega$}}  &
{\textbf{VaR$_{5\%}^{(\text{loss})}$}} & {\textbf{MES$_{5\%}^{(\text{loss})}$}} & {\textbf{$\Omega$}}  &
{\textbf{VaR$_{5\%}^{(\text{loss})}$}} & {\textbf{MES$_{5\%}^{(\text{loss})}$}} & {\textbf{$\Omega$}}  &
{\textbf{VaR$_{5\%}^{(\text{loss})}$}} & {\textbf{MES$_{5\%}^{(\text{loss})}$}} & {\textbf{$\Omega$}}  \\ 
\midrule
\textbf{1} & 3.04 & 3.46 & 4.97 & 3.08 & 2.66 & 3.35 & 2.09 & -4.62 & 3.57 & 1.76 & -4.68 & 4.48 \\
\textbf{2} & 2.20 & 4.13 & 3.28 & 2.96 & 3.82 & 2.27 & 2.14 & -0.23 & 2.77 & 2.49 & -0.29 & 3.64 \\
\textbf{3} & 3.13 & 2.51 & 2.24 & 3.53 & 2.59 & 1.36 & 3.98 & 4.35 & 1.31 & 3.91 & 3.77 & 1.70 \\
\textbf{4} & 2.61 & 3.10 & 2.09 & 2.76 & 2.75 & 1.80 & 4.93 & 4.14 & 1.69 & 4.61 & 3.94 & 2.05 \\
\textbf{5} & 2.16 & 2.96 & 2.17 & 3.10 & 2.33 & 2.47 & 3.49 & 1.95 & 1.49 & 3.24 & 1.86 & 1.96 \\
\textbf{6} & 3.13 & 3.41 & 1.51 & 2.96 & 2.19 & 1.96 & 3.74 & 2.51 & 1.32 & 3.41 & 2.39 & 1.64 \\
\textbf{7} & 2.43 & 2.87 & 1.30 & 2.03 & 1.62 & 1.70 & 3.16 & 1.50 & 1.30 & 3.16 & 1.37 & 1.37 \\
\textbf{8} & 2.39 & 1.81 & 1.61 & 1.67 & 1.94 & 1.94 & 2.94 & 1.62 & 1.49 & 2.87 & 1.53 & 1.53 \\
\textbf{9} & 2.33 & 1.40 & 1.17 & 2.31 & 1.55 & 1.32 & 2.74 & 1.66 & 1.08 & 3.11 & 1.61 & 1.06 \\
\textbf{10} & 2.31 & 1.02 & 1.09 & 1.50 & 0.99 & 1.58 & 2.35 & 1.98 & 0.91 & 1.95 & 1.86 & 0.90 \\
\textbf{11} & 2.15 & 3.78 & 1.03 & 1.63 & 2.78 & 1.39 & 2.20 & 4.64 & 0.77 & 2.19 & 4.70 & 0.73 \\
\textbf{12} & 2.54 & 1.93 & 1.06 & 2.37 & 1.74 & 1.24 & 3.13 & 2.81 & 0.77 & 3.11 & 2.73 & 0.68 \\
\textbf{13} & 2.19 & 1.72 & 1.03 & 2.39 & 1.53 & 1.15 & 3.16 & 2.77 & 0.75 & 3.16 & 2.72 & 0.66 \\
\textbf{14} & 2.08 & 1.99 & 0.92 & 1.74 & 1.43 & 1.32 & 3.21 & 2.76 & 0.65 & 2.51 & 1.87 & 0.67 \\
\bottomrule
\end{tabular}
\caption{%
5\% Value-at-Risk (VaR$_{5\%}$), Marginal Expected Shortfall (MES$_{5\%}$), 
and Omega ratio ($\Omega$) across portfolio strategies and holding periods $\Delta_h$.
VaR and MES are reported as positive loss magnitudes (in \%) computed as
$\text{VaR}_{5\%}^{(\text{loss})} = -\text{VaR}_{5\%}(R_p) \times 100 $ and 
$\text{MES}_{5\%}^{(\text{loss})} = - \mathbb{E}[R_p \mid R_m \le \text{VaR}_{5\%}(R_m)]  \times 100$.
VaR and MES correspond to daily returns, and $\Omega$ ratios are unitless.
The column of $\Omega$ for strategy P(ARIMA) that performed well across all holding periods is shaded in grey.}
\label{tab:VaR_MES_Omega}
\end{table}

Our strategies achieve moderate tail risk (Table~\ref{tab:VaR_MES_Omega}). For very short-term (1–3 days), $\text{VaR}_{5\%}^{(\text{loss})}$ lies between 2-4\%, and $\Omega$ is above unity (e.g., $\Omega\in[3.3,5.0]$ for $\Delta_h=1$). 
As $\Delta_h$ increases, tail risk generally compresses for P(ARIMA) (e.g., $\text{VaR}_{5\%}^{(\text{loss})}$ declines from 3.08\% ($\Delta_h=1$) to 1.74\% ($\Delta_h=14$) and $\text{MES}_{5\%}^{(\text{loss})}$ from 2.66\% to 1.43\%), while $\Omega$ remains above one up to $\Delta_h=14$). The predictive strategy not only preserves profitability on average but also maintains comparatively tighter left-tail exposure at medium to long horizons, in line with its stronger AT, WR, and PF profiles beyond one week. The S and P(ARIMA)–S variants show a progressive deterioration in asymmetry for longer holds: $\Omega$ falls below one from $\Delta_h=10$ onward (e.g., $\Omega\in[0.65,0.91]$ ($\Delta_h\in[10,14]$), with larger $\text{MES}_{5\%}^{(\text{loss})}$ spikes around $\Delta_h=$ 11–13. The Baseline sits between these extremes: its short-horizon \(\Omega\) is strong but drifts below unity by \(\Delta_h=14\), and $\text{VaR}_{5\%}^{(\text{loss})}$ declines less than under P(ARIMA). Overall, the tail-risk evidence reinforces the main findings from our AT, WR, and PF analysis, i.e. predictive information yields more resilient risk–reward at extended horizons, whereas consensus clustering without prediction is comparatively fragile in the tails.

Some negative $\text{MES}_{5\%}^{(\text{loss})}$ values appear in the S and P(ARIMA)–S portfolios at the shortest horizons, suggesting that these portfolios occasionally achieve gains during systemic market downturns. This behaviour reflects the stabilising role of consensus clustering combined with adaptive window shifting, which aligns portfolio composition with transient but more resilient market structures. By rebalancing on the basis of evolving inter-asset correlations rather than static returns, the shifting mechanism can temporarily identify clusters that are partially insulated from market-wide shocks, particularly when local volatility regimes differ across assets. In such settings, these portfolios may exhibit short-lived countercyclical characteristics, generating positive conditional returns when the aggregate market performs poorly. This effect diminishes as the holding horizon increases because cluster persistence weakens, tail co-movement among cryptocurrencies intensifies, and MES reverts to more conventional positive loss levels, signalling increased systemic vulnerability. Thus, while shifted consensus clustering can provide temporary hedging benefits under turbulent yet segmented market conditions, its protective capacity fades as horizons extend and correlations converge under broader market stress.

\subsection{Crypto-market financial context}

Between 2017 and 2022, the cryptocurrency market progressed through several distinct phases, including the speculative boom of 2017–2018, the subsequent correction in 2018–2019, and a more adoption-driven expansion from 2019 onward. External shocks such as the COVID-19 pandemic and episodes of elevated inflation introduced additional structural breaks. These regime shifts generated markedly different volatility and correlation patterns, which in turn affected portfolio performance. 
While our indicators suggest overall profitable performance, realised excess returns vary across regimes. The implied daily returns indicate that the strategies generate sufficient alpha to exceed standard daily inflation rates (historically $<0.03\%$), especially for shorter holding periods. These returns also provide a buffer against typical cryptocurrency trading costs (approximately $0.10\%$–$0.20\%$ per side on major exchanges), such that at shorter horizons the strategies can remain profitable after transaction fees, provided that execution slippage and bid–ask spreads are well managed. The framework therefore demonstrates a statistically and economically meaningful edge before costs, although net performance remains sensitive to fee levels, liquidity conditions, and implementation details. 
In practice, an investor can use our framework to target a desired excess return over the risk-free rate by selecting the strategy and holding horizon that jointly deliver higher AT and $\Omega>1$ with controlled tail risk.

Our consensus clustering frequently identifies Bitcoin (BTC) as part of the largest and most persistent clusters, reflecting its role as a dominant asset and market-wide reference point. This pattern is consistent with herd behaviour in cryptocurrency markets, where shocks to BTC often propagate through correlated assets and undermine na\"ive diversification. By aggregating community partitions over time, our consensus procedure filters out short-lived, sentiment-driven correlations and recovers groups of cryptocurrencies that consistently co-move across regimes. These groups can be interpreted as risk classes, enabling investors to diversify or hedge more effectively by selecting assets across classes rather than treating cryptocurrencies as a homogeneous category. From a network perspective, the strong common market component naturally compresses global modularity, yet the consensus clusters remain sufficiently distinct to enhance diversification and improve tail-risk control in the portfolio experiments.

Behavioural factors further shape these interdependencies. The discrepancies in prediction accuracy during abrupt market changes can be interpreted through investor overreaction to regulatory news, macroeconomic surprises, or sentiment cascades. Such shocks generate temporary deviations from typical co-movement patterns. Historical cluster structures provide a principled basis for anticipating subsequent reversion and for tactical rebalancing when asset prices diverge from their usual correlation-based groupings.

The inclusion of VaR, MES, and the Omega ratio contextualises profitability and vulnerability across strategies. For P(ARIMA) portfolios, declining $\text{VaR}_{5\%}$ and $\text{MES}_{5\%}$ at longer horizons, together with AT $>0$, WR $>0.5$, PF $>1$, and $\Omega>1$, indicate that observed returns are achieved with controlled downside risk and favourable gain–loss asymmetry. By contrast, the deterioration of $\Omega$ and elevated MES for the S and P(ARIMA)–S strategies at extended horizons reveal that, under structural stress, their raw returns are less likely to translate into attractive risk-adjusted performance. These results show that stable clustering structures influence not only co-movement but also the concentration of downside risk.

To anchor these findings in an asset-pricing perspective, we adopt a risk-free rate $r_f=0.02$, consistent with long-term yields on stable government securities. 
Given that our returns are reported on an implied daily scale, this annual benchmark translates to a negligible daily hurdle ($\approx 0.005\%$). Consequently, the positive return values observed across most horizons confirm that the portfolios generate true economic profits beyond the opportunity cost of capital.
This benchmark does not affect realised portfolio returns, which are generated purely by price appreciation, but provides a disciplined reference for evaluating excess returns and risk-adjusted metrics. 
Interpreting profitability and tail-risk measures relative to $r_f$ clarifies each strategy’s capacity to generate persistent excess returns. Taken together, our experiments indicate that predictive clustering strategies exhibit a stronger ability to maintain profitability, control downside exposure, and deliver economically meaningful excess returns in a structurally dynamic and behaviourally influenced cryptocurrency market.

\section{Conclusion}

This study introduces an integrated framework that combines predictive analytics, network-based clustering, and modern portfolio theory to enhance cryptocurrency selection and diversification in designing optimal portfolios. Central to our methodology is the identification of persistent clusters of highly interdependent cryptocurrencies via consensus clustering. These clusters capture structural co-movement patterns that remain stable across market regimes, enabling portfolios to diversify across risk classes rather than relying on na\"ive asset-level diversification, where correlations can rise sharply during stress.

By incorporating predicted returns into the correlation network, the proposed P(ARIMA) strategy improves the detection of forward-looking interdependencies and provides better-performing portfolios over extended holding horizons. Although predictive models are computationally demanding, our empirical findings show that prediction-enhanced clustering systematically strengthens gain–loss asymmetry, reduces tail-risk exposure, and maintains profitability as horizons increase. In contrast, strategies relying purely on historical information or window shifting exhibit weaker performance persistence, with risk–reward profiles deteriorating more rapidly when horizons lengthen, or correlations tighten. These results highlight that predictive information is most valuable not for direct price estimation, but for refining the identification of stable correlation structures that support effective diversification. On the other hand, the high computational costs of forecasting models may not justify their adoption compared to our simpler strategies.

Future research should explore applying our methodology to higher-frequency data to capture intra-day dynamics, incorporate trading frictions explicitly, evaluate alternative risk measures, and integrate cryptocurrencies with other asset classes. The clustering step may be improved by using more advanced methods of noise removal, and future work could examine the use of return variability or volatility-adjusted similarity measures when estimating inter-asset dependencies. Network-based clustering could also be complemented with feature- or behaviour-based machine learning methods to test whether hybrid approaches yield additional gains in stability or interpretability. Nonetheless, the network-based consensus methodology used here retains important advantages, such as interpretable clusters, reduced sensitivity to stochastic variability, and scalability to larger cryptocurrency universes.

Overall, our results demonstrate that cryptocurrencies exhibit sufficiently persistent interdependencies to support structured portfolio design and systematic diversification. Network-based stable clustering strategies offer principled and adaptive means of exploiting these interdependencies, enabling investors to construct portfolios that manage risk effectively while capturing opportunities across regimes. As cryptocurrency markets mature, institutional investors enter the market, and regulatory frameworks evolve, methodologies that integrate prediction, clustering, and risk-aware optimisation are increasingly valuable for investors seeking robust performance in a highly volatile and rapidly changing financial landscape.

\clearpage

\section*{Acknowledgements}
R.J. is funded by the China Scholarship Council (CSC) from the Ministry of Education of P. R. China. R.K. is partially funded by JSPS KAKENHI (Nos. JP18K11560, JP21H03559, JP21H04571, JP22H03695, and JP23K24950), JST PRESTO (No. JPMJPR1925), and AMED (No. JP223fa627001). L.E.C.R. is partially funded by the FWO Scientific Research Network (W001625N), and the Bijzonder Onderzoeksfonds (BOF/STA/201909/022) from Ghent University, Belgium.

\bibliography{maintext}

\setcounter{figure}{0}
\setcounter{table}{0}
\renewcommand{\thefigure}{S\arabic{figure}}
\renewcommand{\thetable}{S\arabic{table}}

\section*{Supplementary Information}

\subsection*{Data source}

The daily price data of the cryptocurrencies used in our study was obtained from the following websites:
www.investing.com, coinmarketcap.com, www.coindesk.com, www.coincodex.com and www.marketwatch.com. Table~\ref{Tab:currency} shows the code of all the cryptocurrencies used in our study.

\begin{table}[htp]
    \centering
    \small
    \begin{tabular}{llllllllll}
    \hline
        ADA & BLOCK & OCEANp & PPC & ETH & QTUM & STORJ & LBC & MTL & XCP \\ 
        ADX & BNB & OCN & CND & ETP & RCN & STRAX & LINK & MYST & XDN \\ 
        AE & BNT & OK & CVC & EVX & RDD & SWFTC & LRC & NAS & XEM \\
        AION & BTC & OMG & DASH & FAIR & RDN & SYNX & LSK & TOA & XLM \\
        AMB & BTM & OMNI & DCN & FCT & REV & SYS & LTC & TRX & XRP \\
        ANT & BTS & ONION & DCR & FLASH & RLC & GBYTE & LUNA & TUBE & XST \\ 
        AOAR & BTU & PAC & DENT & FLO & SALT & GEO & MAID & UBQ & XTZ \\ 
        ARDR & NEBL & PART & DGB & FTC & SBDR & GLM & MANA & USDT & XVG \\ 
        ARK & NEO & PASC & DIME & FUEL & SC & GRC & MDA & VAL & ZAP \\ 
        AST & NEU & PAY & DLT & FUN & SCRT & ICX & MDT & VERI & ZCL \\ 
        BAT & NLG & PHR & DNT & GAME & SMART & IGNIS & MHC & VET & ZEC \\ 
        BCD & NMC & PHX & DOGE & GAS & SNC & IOC & IOTA & VIA & ZEN \\
        BCH & NMR & PIVX & EDG & PPT & SNM & JNT & MKR & VIB & ZRX  \\ 
        BCN & NXS & PLR & EMC2 & PRE & SNT & KEY & MLN & WAVES & ~ \\
        BITCNY & NXT & PND & ENG & PRO & SOUL & KMD & MONA & WGR & ~ \\ 
        BLK  & OAX  & POT  & ETC  & QRL  & STEEM  & KNC  & MTH  & WINGS  & ~ \\ \hline
    \end{tabular}
    \caption{The list of the $N=157$ cryptocurrencies used in our study.}
    \label{Tab:currency}
\end{table}

\subsection*{Portfolio performance with various correlation network threshold}

To assess the robustness of the portfolio construction strategies, we examine how varying the correlation threshold $\theta_{\rho}$ used in building the cryptocurrency correlation network affects portfolio performance. Modifying $\theta_{\rho}$ alters the sparsity and modular structure of the correlation network, where lower thresholds include weaker links resulting in larger, denser communities, while higher thresholds isolate only the strongest dependencies. For each threshold ($\theta_{\rho} \in \{0.4, 0.5, 0.6\}$), portfolios were constructed according to the four strategies, and the corresponding Average Trade (AT) values were computed across investment horizons.

\begin{figure}[h]
\raggedright
    \includegraphics[scale=0.35]{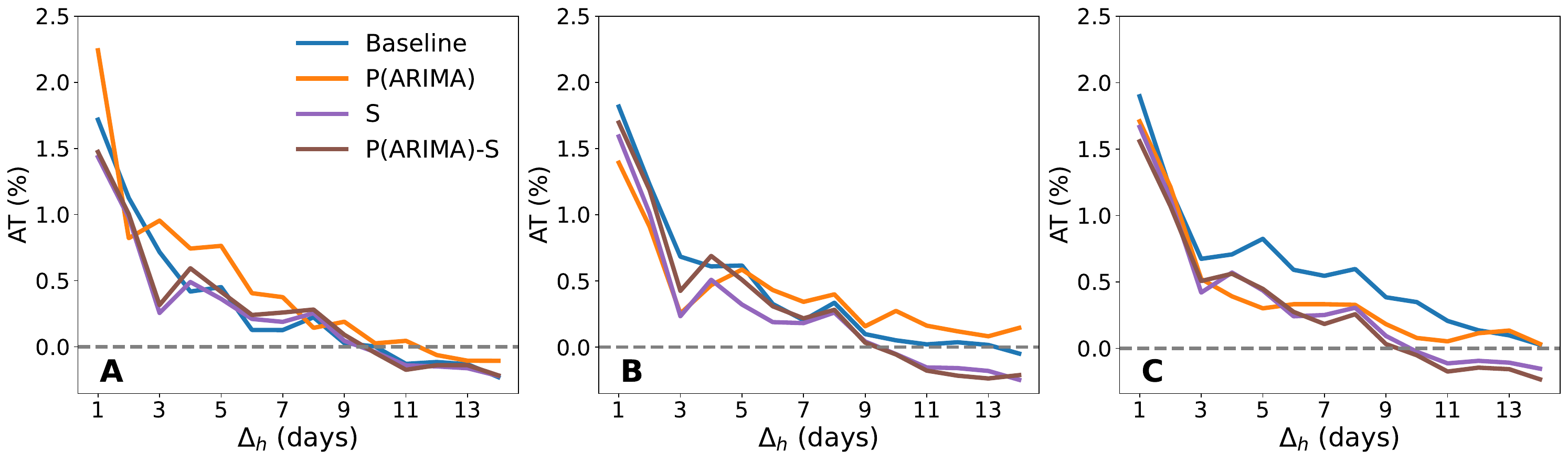}
    \caption{Average Trade (AT) performance for four portfolio construction strategies Baseline, P(ARIMA), S, and P(ARIMA)-S, across investment horizons ($\Delta_h$ = 1–14 days) under three correlation thresholds used for the correlation network: (A) $\theta_{\rho}=0.4$, (B) $\theta_{\rho}=0.5$, and (C) $\theta_{\rho}=0.6$. The dashed horizontal lines indicate the reference level of $AT=0$. Returns are measured on an implied daily simple scale.}
    \label{fig:threshold_portfolios}
\end{figure}

As illustrated in Fig.~\ref{fig:threshold_portfolios}, the portfolio performance profiles exhibit stability across all thresholds, and strategies consistently deliver positive initial returns up to 9-day holding periods, with 1st day values remaining well above $1.3\%$ in every scenario. This confirms that the profitability of the consensus clusters is not an artefact of fine-tuning a specific threshold but rather a result of capturing persistent structural dependencies.
In the meantime, a shift in relative dominance is observable as the network topology changes. In the denser network ($\theta_{\rho}=0.4$, Fig.~\ref{fig:threshold_portfolios}A), the P(ARIMA) strategy outperforms the Baseline, suggesting that predictive information effectively filters noise when weaker correlations are present. Conversely, in the sparser network ($\theta_{\rho}=0.6$, Fig.~\ref{fig:threshold_portfolios}C), where only the strongest links remain, the Baseline strategy assumes a slight lead. This indicates that while the strategies are robust overall, P(ARIMA) offers advantages in complex, high-connectivity environments, whereas the Baseline suffices when the correlation structure is naturally sparse and clean.

\subsection*{Alternative benchmark portfolio strategies}

To benchmark the performance of our proposed strategies, we further compare them against two alternative portfolio construction strategies: a single-asset Bitcoin strategy (BTC) and a low-volatility selection (Vol10). The BTC strategy holds Bitcoin only and serves as a single-asset baseline representing the market leader. The Vol10 strategy selects the 10 cryptocurrencies with the lowest volatility for each study period, and weights are optimised using the same mean–variance procedure as in our proposed strategies.

\begin{figure}[h]
\raggedright
    \includegraphics[scale=0.33]{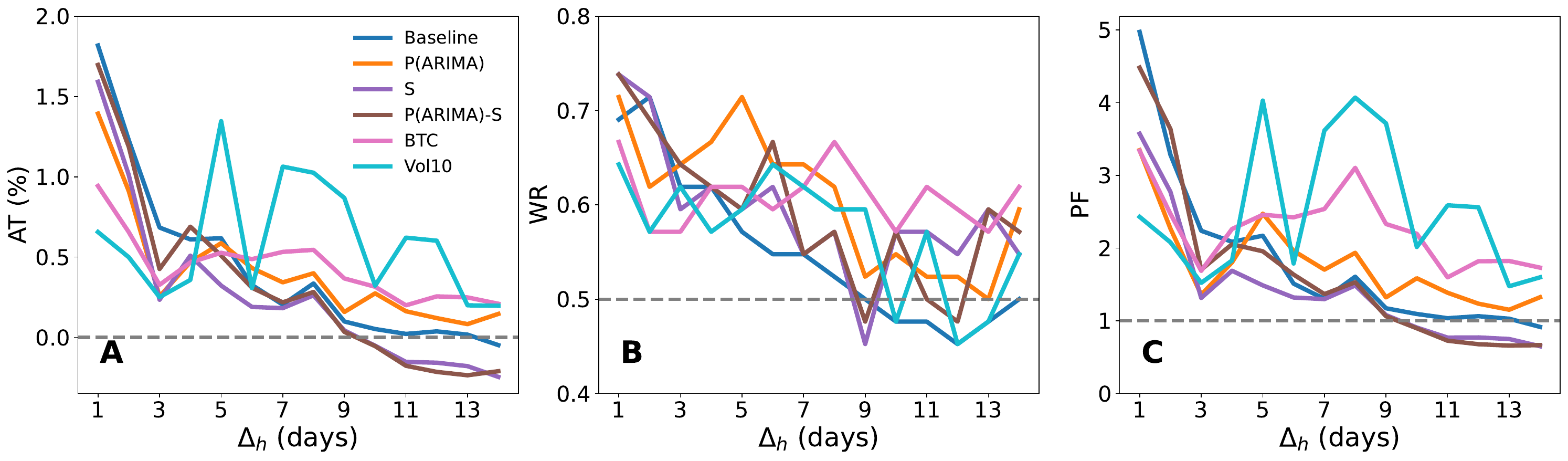}
    \caption{Performance indicators of the proposed strategies for Baseline, P(ARIMA), S, and P(ARIMA)–S with alternative portfolio constructions, Bitcoin only (BTC), and low-volatility selection (Vol10). Shown are (A) Average Trade (AT), (B) Win Rate (WR), and (C) Profit Factor (PF) across investment horizons $\Delta_h = 1$-14 days. The dashed horizontal lines indicate the reference levels of $AT = 0$, $WR = 0.5$, and $PF = 1$, respectively. Returns are measured on an implied daily simple scale.}
    \label{fig:benchmark_indicators}
\end{figure}

Figure~\ref{fig:benchmark_indicators} highlights the distinct performance characteristics of the consensus-based strategies relative to these heuristics. In the immediate short term, our proposed strategies exhibit a decisive advantage, achieving significantly higher AT and PF values than both benchmarks. 
However, the dynamics shift as the investment horizon lengthens. The Vol10 strategy, while unstable, occasionally surges to achieve the highest performance peaks, indicating that naive low-volatility selection can capture intermittent upside despite its lack of consistency.
Furthermore, the single-asset BTC benchmark demonstrates superior resilience at extended horizons, eventually outperforming the proposed strategies. However, relying solely on BTC entails significant idiosyncratic risk due to a complete lack of diversification, which results in a lower PF compared to the optimised portfolios during the initial windows. This suggests a temporal trade-off, that the proposed network-based strategies are highly effective at exploiting short-term structural alpha, but as the correlation structure evolves over time, the advantages of complex diversification fade, eventually yielding to the simpler beta performance of the market leader.

\subsection*{Network modularity under varying correlation thresholds}

To assess the sensitivity of the detected community structures to the choice of correlation threshold, we evaluate how the average modularity $\langle Q \rangle$ and its variability change as $\theta_{\rho}$ varies from 0.1 to 0.9 across all $n_s = 42$ study periods. For each strategy, modularity is computed on the corresponding thresholded correlation networks, and the resulting distribution over study periods is summarised through both its mean value and standard deviation.

\begin{figure}[h]
    \centering
    \includegraphics[width=0.24\textwidth]{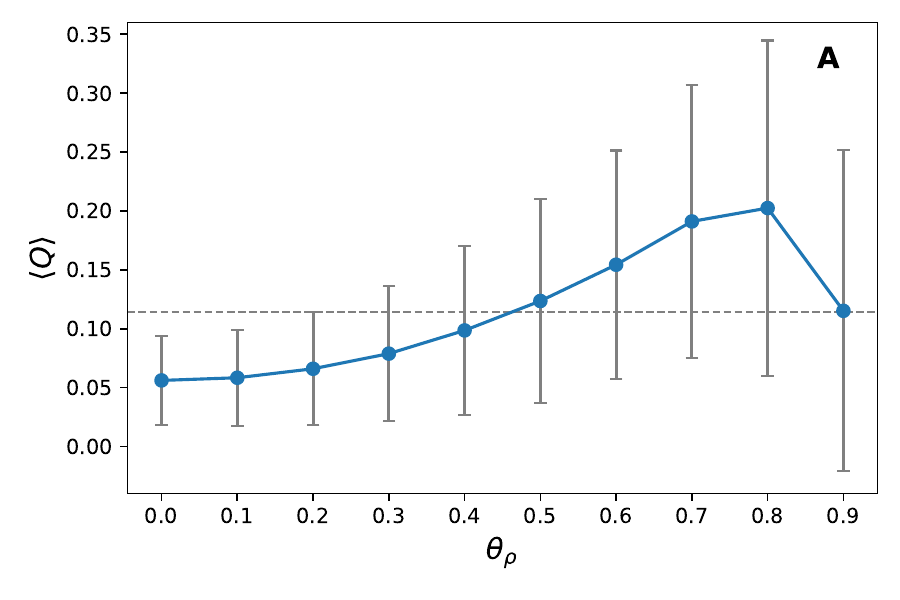}
    \includegraphics[width=0.24\textwidth]{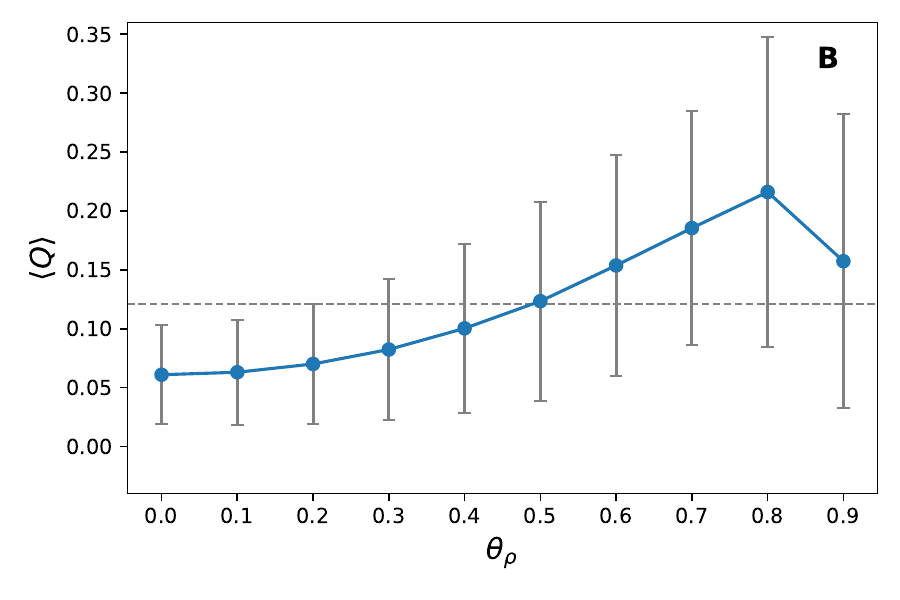}
    \includegraphics[width=0.24\textwidth]{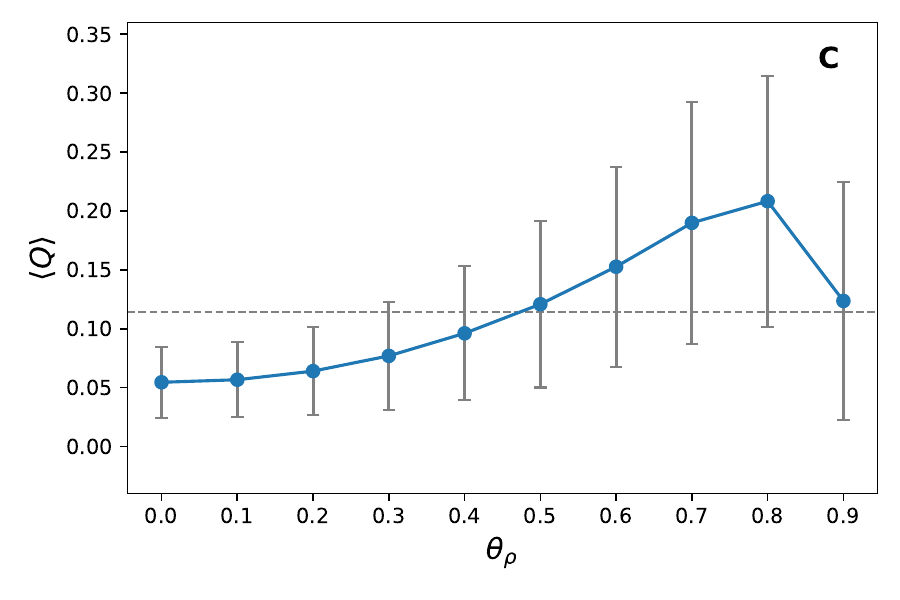}
    \includegraphics[width=0.24\textwidth]{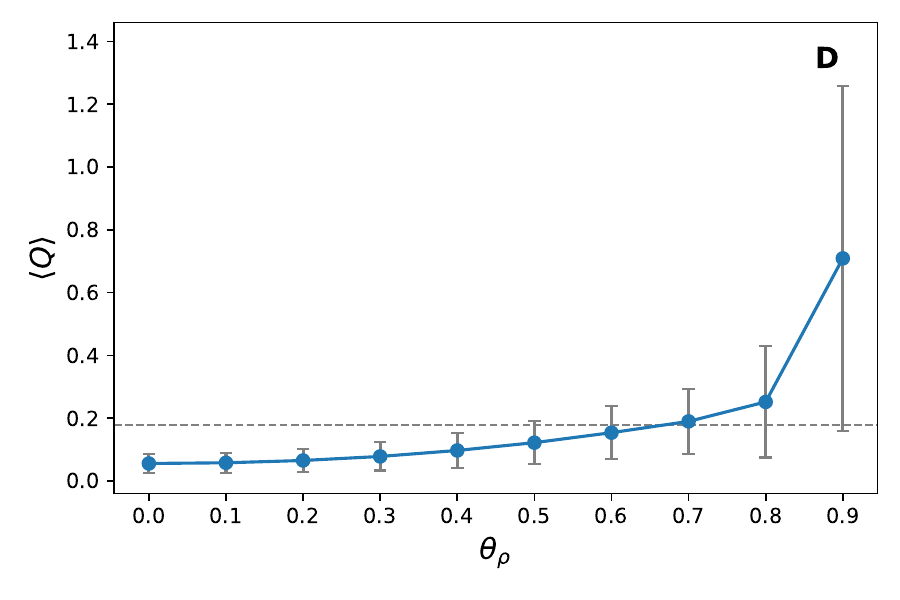}
    \caption{
    Average modularity $\langle Q \rangle$ and corresponding standard deviation $\sigma$ across all $n_s = 42$ study periods for correlation thresholds $\theta_{\rho}$ ranging from 0.1 to 0.9, for strategies (A) Baseline, (B) P(ARIMA), (C) S, and (D) P(ARIMA)-S. Vertical bars represent the standard deviation of modularity across periods. The horizontal dashed line denotes the mean modularity value averaged over all thresholds. }
    \label{fig:threshold_curves}
\end{figure}

\begin{figure}[h]
    \centering
    \includegraphics[width=0.24\textwidth]{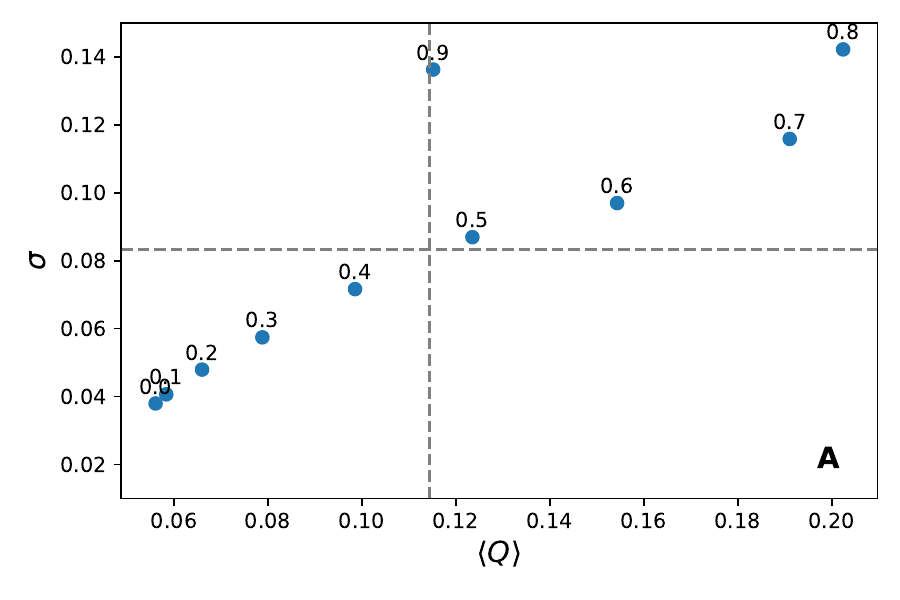}
    \includegraphics[width=0.24\textwidth]{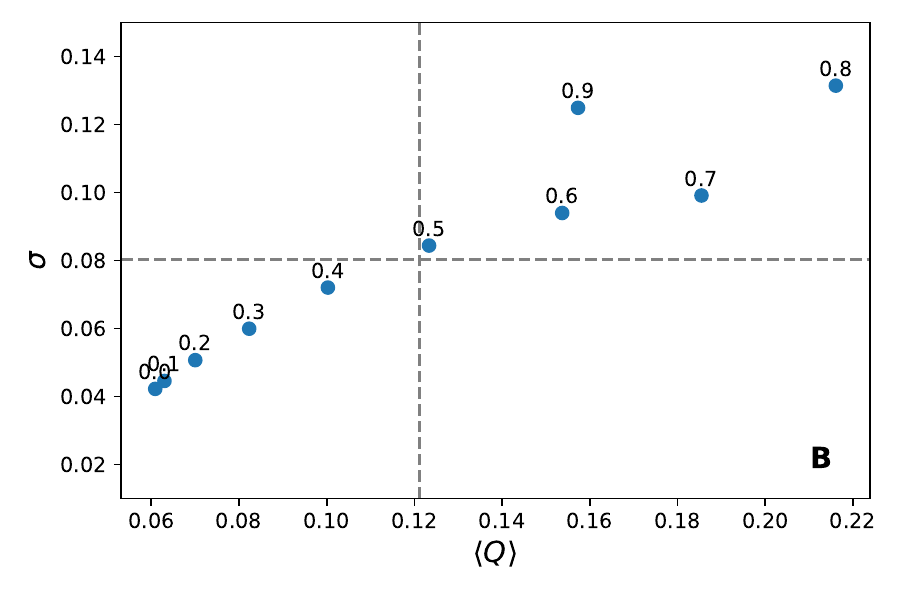}
    \includegraphics[width=0.24\textwidth]{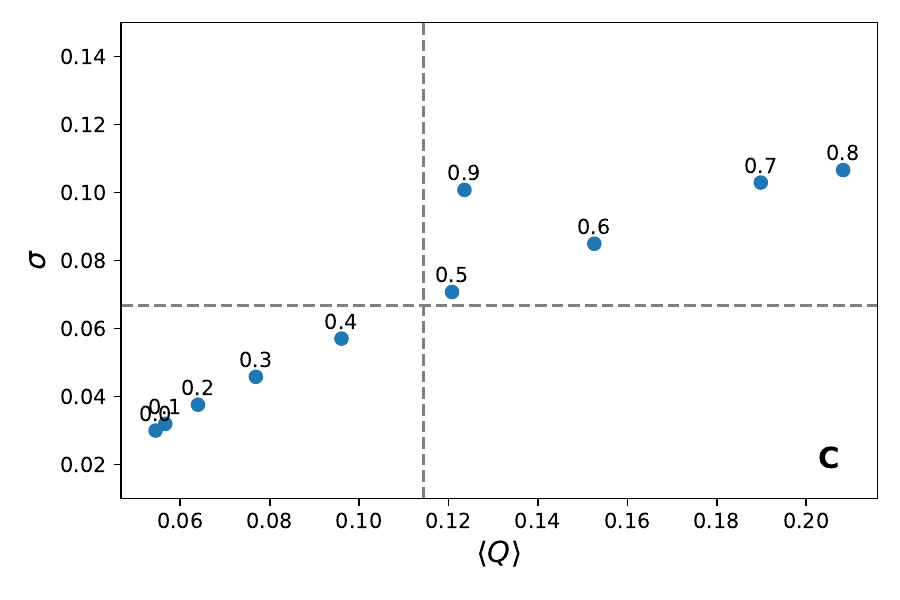}
    \includegraphics[width=0.24\textwidth]{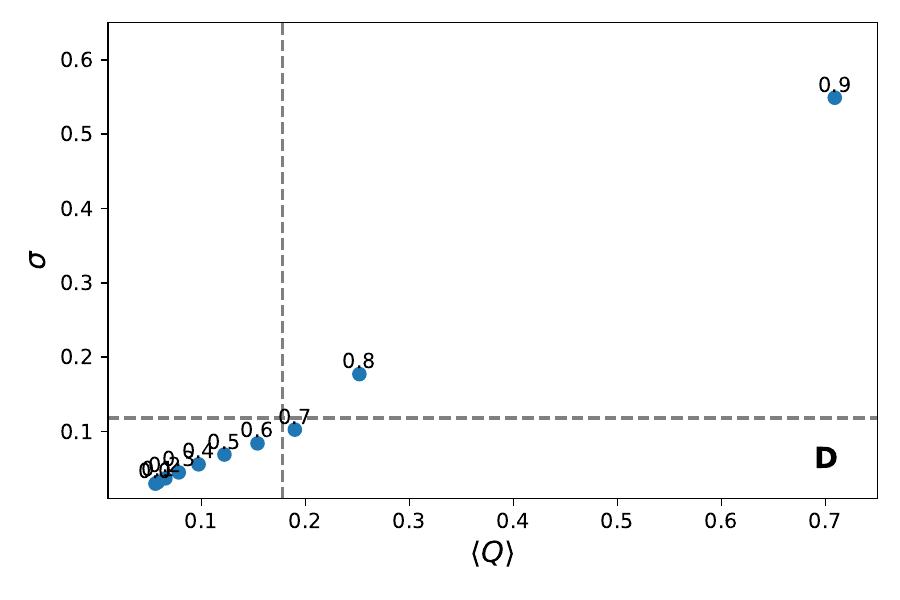}
    \caption{Trade-off between average modularity $\langle Q \rangle$ (x-axis) and its standard deviation $\sigma$ (y-axis) over all $n_s = 42$ study periods for strategies (A) Baseline, (B) P(ARIMA), (C) S, and (D) P(ARIMA)-S. Each point corresponds to a correlation threshold $\theta_{\rho}$, with thresholds sampled from 0.1 to 0.9. The horizontal dashed line marks the overall mean $\langle Q \rangle$ across thresholds, and the vertical dashed line marks the overall mean $\sigma$. }
    \label{fig:threshold_tradeoffs}
\end{figure}

Figure~\ref{fig:threshold_curves} shows that, for all strategies, modularity tends to increase as the correlation threshold becomes more restrictive, reflecting the emergence of more clearly separated clusters when weaker correlations are pruned. However, high thresholds can also induce instability, as indicated by the increases in standard deviation. 
To evaluate the trade-off between cluster strength and stability, Fig.~\ref{fig:threshold_tradeoffs} displays $\langle Q \rangle$ against its standard deviation $\sigma$ for each threshold level. Across all strategies, $\theta_{\rho} = 0.5$ lies close to the intersection of the mean modularity and mean standard deviation reference lines, indicating that 0.5 provides a balanced trade-off between cluster separability and stability across periods. 
These results support the choice of $\theta_{\rho} = 0.5$ in the main text, as it provides a representative middle ground that avoids the noisy behaviour of very low thresholds and the excessive fragmentation and instability associated with very high thresholds.

\end{document}